\documentclass[aps,prx,twocolumn,superscriptaddress,longbibliography]{revtex4-2}

\usepackage{amsmath}
\usepackage{amssymb}
\usepackage{amsfonts}
\usepackage{bm}
\usepackage{graphicx}

\usepackage{braket}

\usepackage[unicode=true,
 bookmarks=false,
 breaklinks=false,pdfborder={0 0 1},backref=false,colorlinks=true]
 {hyperref}
\hypersetup{
 linkcolor=blue,citecolor=blue,urlcolor=blue,linkcolor=blue,citecolor=blue,urlcolor=blue}
\usepackage{microtype}

\makeatletter

\DeclareMathOperator{\tr}{\mathrm{Tr}}

\newcommand{\cqp}{\affiliation{Center for Quantum Physics, University of Innsbruck, 6020 Innsbruck, Austria}}
\newcommand{\iqoqi}{\affiliation{Institute for Quantum Optics and Quantum Information of the Austrian Academy of Sciences,  6020 Innsbruck, Austria}}
\newcommand{\aei}{\affiliation{Institute for Theoretical Physics, Institute for Gravitational Physics (Albert Einstein Institute), Leibniz University Hannover, 30167 Hannover, Germany}}

\begin{abstract}
We discuss quantum variational optimization of Ramsey interferometry with ensembles of $N$ entangled atoms, and its application to atomic clocks based on a Bayesian approach to phase estimation.
We identify best input states and generalized measurements within a variational approximation for the corresponding entangling and decoding quantum circuits.
These circuits are built from basic quantum operations available for the particular sensor platform, such as one-axis twisting, or finite range interactions.
Optimization is defined relative to a cost function, which in the present study is the Bayesian mean squared error of the estimated phase for a given prior distribution, i.e.~we optimize for a finite dynamic range of the interferometer.
In analogous variational optimizations of optical atomic clocks, we use the Allan deviation for a given Ramsey interrogation time as the relevant cost function for the long-term instability.
Remarkably, even low-depth quantum circuits yield excellent results that closely approach the fundamental quantum limits for optimal Ramsey interferometry and atomic clocks.
The quantum metrological schemes identified here are readily applicable to atomic clocks based on optical lattices, tweezer arrays, or trapped ions. While in the present work variationally optimized circuits are found with classical simulations, optimization can also be performed `on' the (physical) quantum sensor, also in regimes not accessible to classical computations and in presence of imperfections. 
\end{abstract}

\begin{document}

\title{Quantum Variational Optimization of Ramsey Interferometry and Atomic Clocks}

\author{Raphael Kaubruegger${}^*$}\cqp \iqoqi
\thanks{Equal author contribution}
\author{Denis V. Vasilyev${}^*$}\cqp \iqoqi
\thanks{Equal author contribution}
\author{Marius Schulte}\aei
\author{Klemens Hammerer}\aei
\author{Peter Zoller}\cqp \iqoqi

\maketitle

\section{Introduction}

Recent progress in quantum technology of sensors has provided us with the most precise measurement devices available in physical sciences. Examples include the development of optical clocks~\cite{Ludlow2015}, atom~\cite{Cronin2009} and light~\cite{Demkowicz2015} interferometers,  and magnetic field sensing~\cite{Degen2017}. These achievements have opened the door to novel applications from the practical to the scientific. Atomic clocks and atomic interferometers allow height measurements in relativistic geodesy~\cite{Grotti2018, Mehlstaubler2018, Bothwell2021, Zheng2021} or fundamental tests
of our understanding of the laws of nature~\cite{Kolkowitz2016, Safronova2018, Sanner2019}, such as time variation of the fine structure constant. 
In the continuing effort to push the boundaries of quantum sensing, entanglement as a key element of quantum physics gives the opportunity to reduce quantum fluctuations inherent in quantum measurements below the standard quantum limit (SQL), i.e. what is possible with uncorrelated constituents~\cite{Pezze2018}. 
Squeezed light improves gravitational wave detection~\cite{Tse2019}, allows lifescience microscopy below the photodamage limit~\cite{Casacio2021}, further, squeezing has been demonstrated in atom interferometers~\cite{Leibfried2005, Appel2009, Gross2010, Riedel2010, Leroux2010, Lucke2011, Sewell2012, Hamley2012, Berrada2013, Barontini2015, Zhang2015, Bohnet2016, Hosten2016a, Norcia2018, Zou2018, PedrozoPeafiel2020}. However, beyond the SQL, quantum physics imposes ultimate limits on quantum sensing, and one of the key challenges is to identify, and in particular devise experimentally realistic strategies defining \emph{optimal quantum sensors}~\cite{Macieszczak2014}.

\begin{figure*}[t]
    \centering
    \includegraphics[]{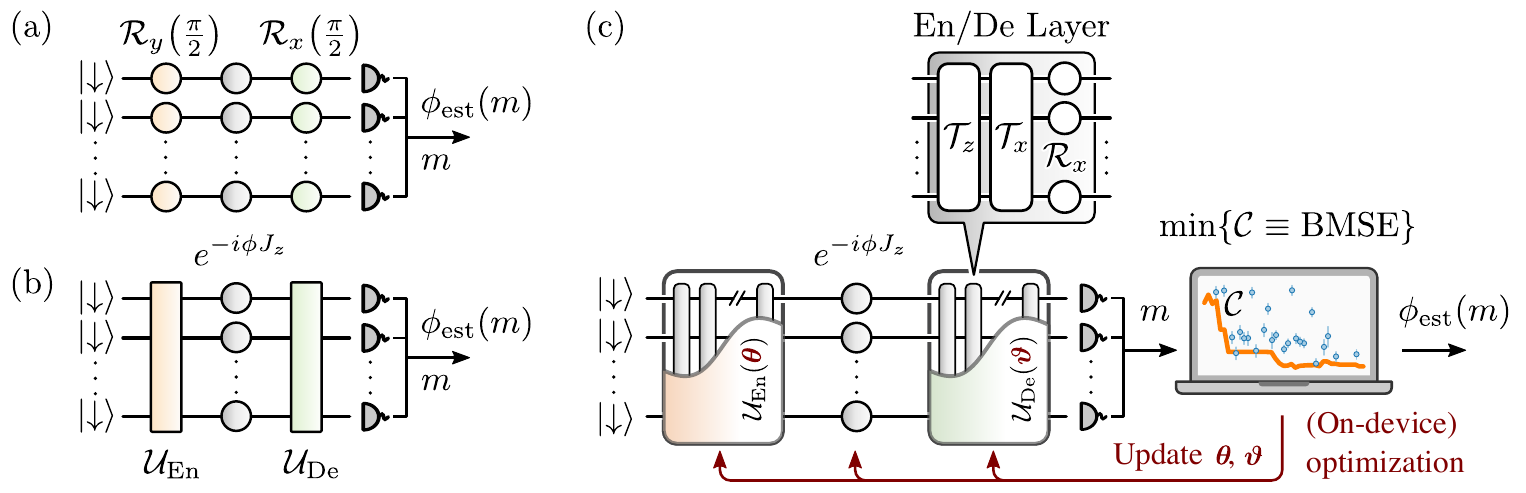}
    \caption{(a) Quantum circuit representation of Ramsey interferometer with uncorrelated atoms. The phase $\phi$ is imprinted on the atomic spin-superposition prepared by global $\pi/2$-rotation around $y$-axis, $\mathcal{R}_y(\pi/2)$.  Consequent rotation, $\mathcal{R}_x(\pi/2)$, and measurement of difference $m$ of atoms in eigenstates $\ket{\uparrow}$ and $\ket{\downarrow}$ in $z$-basis allows estimating the phase $\phi$ using an estimator function $\phi_{\rm est}(m)$.
    (b) Quantum circuit of a generalized Ramsey interferometer with generic \emph{entangling} and \emph{decoding} operations $\mathcal{U}_{\rm En}$ and $\mathcal{U}_{\rm De}$, respectively. 
    Our variational approach (c) consists of an ansatz, where optimal $\mathcal{U}_{\rm En}$ and $\mathcal{U}_{\rm De}$ are approximated by low-depth circuits. These are built from `layers' of elementary operations, which are provided by the given platform. We specify the variationally optimized quantum sensor by circuits $\mathcal{U}_{\rm En}(\boldsymbol{\theta})$ and $\mathcal{U}_{\rm De}(\boldsymbol{\vartheta})$ [see Eqs.~\eqref{eq:Entangler} and \eqref{eq:Decoder}], of depth $n_{\rm En}$ and $n_{\rm De}$, respectively. Here  $\boldsymbol{\theta}\equiv \{\theta_i\} $ and $\boldsymbol{\vartheta}\equiv \{\vartheta_i\} $ are vectors of variational parameters to be optimized for a given strategy represented by a cost function $\mathcal{C}$ defined here as Bayesian mean squared error (BMSE) [see Eqs.~\eqref{eq:BMSQE_POVM} and \eqref{eq:BMSQE}]. We illustrate the approach with a variational circuit built from global spin rotations $\mathcal{R}_x$ and one-axis-twisting gates $\mathcal{T}_{x,z}$ available in neutral atom and ion quantum simulation platforms, as discussed in Sec.~\ref{subsec:VariationalInterferometer}. The circuits optimization, shown as a feedback loop (in red), can be performed on a classical computer, or, if the complexity of underlying quantum many-body problem exceeds capabilities of classical computers, on the sensor itself, thus, leading to a (relevant) quantum advantage, see Sec.~\ref{sec:FiniteRange}.}
    \label{fig:Fig1}
\end{figure*}

In our discussion below we will focus on optimal Ramsey interferometry, where `optimality' is defined as achieving the best average signal to noise ratio (SNR) for phase estimation in a single-shot measurement. The distinguishing feature of the present work is that we consider optimal Ramsey interferometry with \textit{finite dynamic range}, i.e.~we wish to achieve optimal sensitivity for phases $\phi$ in a given finite interval of width $\delta\phi$~\cite{Personick71, Holevo1982, Buzek1999, Macieszczak2014,Jarzyna2015, Martinez-Vargas2017, Gorecki2020} as is relevant for numerous applications including atomic clocks~\cite{Bloom2014, Ushijima2015, McGrew2018, Origlia2018,  Oelker2019, Chou2010, Huntemann2016, Brewer2019, Covey2019, Wilson2019, Young2020}. To implement this optimal Ramsey interferometry we devise an approach based on variational quantum circuits~\cite{Farhi2014, Wecker2015, McClean2016,OMalley2016, Kokail2019, McArdle2020, Zhou2020, Cerezo2021}. Here, entangled input states and the entangled measurement protocols~\cite{Hosten2016b, Davis2016,Frowis2016,Macri2016,Nolan2017,Anders2018,Schulte2020a} defining the generalized Ramsey interferometer are represented as variational quantum circuits built from `natural' quantum resources available on the specific sensor platform (see Fig.~\ref{fig:Fig1}), which are optimized in light of a given cost function defining the optimal interferometer. 
As we will show, already low-depth variational quantum circuits can provide excellent approximations to the optimal interferometer. Intermediate scale atomic quantum devices~\cite{Deutsch2020, NAP2020}, acting as programmable quantum sensors~\cite{Kaubruegger2019}, present the opportunity to implement these low-depth quantum circuits, defining an experimental route towards \emph{optimal} Ramsey interferometry. 

As noted above, optimality of a quantum sensing protocol is defined via a cost function $\mathcal{C}$ which is identified in context of a specific metrological task. In our study of variational $N$-atom Ramsey interferometry, we wish to optimize for phase estimation accuracy defined as the mean squared error $\epsilon(\phi)$ relative to the actual phase $\phi$, averaged with respect to a  prior distribution $\mathcal{P}(\phi)$ with width  $\delta \phi$, which represents the finite \emph{finite dynamic range} of the interferometer. Thus the cost function is ${\cal C}\equiv (\Delta \phi)^2 = \int d \phi\,  \epsilon(\phi)\mathcal{P}(\phi)$. This corresponds to a Bayesian approach to optimal interferometry where the prior width of the phase distribution $\delta\phi$ is updated through measurement to  $\Delta \phi$ characterizing the posterior distribution. As outlined in Fig.~\ref{fig:Fig1}, our variational approach to optimal Ramsey interferometry seeks to minimize $\cal C$ over variational quantum circuits, and thus identifying optimal input states and measurements for a given~$\delta \phi$. Note that in the present work we optimize a metrological cost function for the complete quantum sensing protocol with variational quantum circuits. We distinguish this from variational state preparation schemes, e.g.~variational squeezed state preparation of Ref.~\cite{Kaubruegger2019}, where a squeezing parameter was optimized as cost function. 

We contrast our Bayesian approach of identifying a metrological cost function to a Fisher information approach, which optimizes accuracy locally at a \textit{specific} value of the phase, corresponding to the limit $\delta\phi\rightarrow 0$~\cite{Demkowicz2015}.  Discussions of fundamental limits in quantum sensing are often phrased in terms of quantum Fisher information and the quantum Cram\`er-Rao bound leading to definition of the Heisenberg limit (HL)~\cite{Braunstein1996, Giovannetti2006, Paris2009}. This  identifies GHZ states~\cite{Bollinger1996}, saturating the HL, as the optimal states for Ramsey interferometry. Furthermore, this leads to the conclusion that adding a decoding step (see Fig.~\ref{fig:Fig1}) is not beneficial for quantum metrology since a separable measurement is optimal in this context~\cite{Giovannetti2006}. This conclusion, however, is not applicable to phase estimation with \emph{finite} prior width since the GHZ state interferometry in single-shot scenarios is optimal only for estimation of phase values in an interval $\delta\phi_{\rm GHZ}\sim1/N$, which shrinks as number of atoms $N$ increases~\cite{Hayashi_2018,Gorecki2020}, see Sec.~\ref{subsec:HL} below. In fact, for large priors $\delta\phi$ tailored quantum input states will differ greatly from squeezed spin states (SSS)~\cite{Wineland1992, Kitagawa1993} or GHZ states~\cite{Macieszczak2014, Demkowicz2015}, and a nontrivial measurement is required for an optimal metrological protocol.  Our variational approach to optimal Ramsey interferometry (see Fig.~\ref{fig:Fig1})  finds these optimal entangling and decoding circuits~\footnote{Variational algorithms for quantum metrology optimizing local phase sensitivity, i.e. maximizing the quantum Fisher information, are considered in Refs.~\cite{Koczor2020, Beckey2020, Ma2020}}.

Our discussion of optimal single-shot Ramsey interferometry~\footnote{While we consider optimal non-adaptive measurement protocols, adaptive protocols for phase estimation have been studied in~\cite{Rosenband2013, Borregaard2013, Kessler2014,Mullan2014, Chabuda2016, Hume2016, Demkowicz2017, Pezze2020a, Pezze2021}.} has immediate relevance for atomic clocks~\cite{Andre2004, Fraas_2016, Leroux2017, Pezze2018, Schulte2020b}. An optical atomic clock operates by locking the frequency of an oscillator,
represented by a classical laser field with fluctuating frequency $\omega_{L}(t)$,
to the transition frequency $\omega_{A}$ of an ensemble of $N$ isolated
atoms~\cite{Ludlow2015}.
The locking of the laser to
the atomic transition is achieved by repeatedly measuring the accumulated
phase $\phi=\int_{0}^{T }dt[\omega_L (t)-\omega_A ]$
in Ramsey interferometry with interrogation time $T$. Importantly, the width $\delta \phi$ of the distribution of this phase increases with the Ramsey time $T$. It is therefore critical to achieve a good phase estimate in conjunction with a wide dynamic range for making an accurate inference about the frequency deviation, and ultimately for stabilizing the clock laser to the atomic transition. Our variational approach for Bayesian phase estimation is made to satisfy these requirements, and provides optimal quantum states and measurements minimizing the instability in atomic clocks as measured by its Allan deviation. We predict significant improvements over previously known one-shot non-adaptive strategies. Our predictions are backed up by comprehensive numerical simulations of the clock laser and its stabilization to the atomic reference in a closed feedback loop~\cite{Leroux2017, Schulte2020b}.

In the following, we first develop the general theory of variationally optimized Ramsey interferometry based on Bayesian phase estimation in Sec.~\ref{sec:interferometer}, and then apply this theory to the specific problem of an optical atomic clock in Sec.~\ref{sec:clock}. 

\section{Quantum Variational Optimization of Ramsey Interferometry}\label{sec:interferometer}

For concreteness, we consider estimation of the phase $\phi$ in an atomic interferometer consisting of an ensemble of $N$ identical two-level atoms described as spin-1/2 particles~\cite{Pezze2018}. The general idea developed in the following applies to any $SU(2)$ interferometer. The interferometer encodes the phase in the atomic state by evolving according to $\ket{\psi_{\phi}}=\exp(-i\phi J_{z})\ket{\psi_{\rm in}}$. Here $\ket{\psi_{\rm in}}$ is an initial probe state \footnote{For simplicity of notation we assume pure states. This discussion is readily extended to mixed states with density operator $\rho_{\rm in}$.}, and ${J}_{x,y,z}=1/2\sum_{k=1}^{N}{\sigma}^{x,y,z}_k$ is the collective spin with $\sigma^{x,y,z}$ the Pauli operators. The task is to determine the unknown phase $\phi$ by performing a measurement on the atoms. 

\subsection{Bayesian approach to phase interferometry}

The most general measurement is described by a positive operator valued measure (POVM), that is a set $\{\Pi_x\}$ of positive Hermitian operators such that $\int dx\,\Pi_x=\openone$. The parameter $\phi$ is estimated on the basis of a measurement result $x$ using an estimator function $\phi_{\rm est}(x)$. The phase estimation accuracy is characterized by a mean squared error (MSE) with respect to the actual phase $\phi$
\begin{equation}
    \epsilon(\phi) =\int dx \big[\phi-\phi_{\rm est}(x)\big]^2 p(x|\phi),
    \label{eq:estimator_variance_POVM}
\end{equation}
where $p(x|\phi)=\tr\{\Pi_x\ket{\psi_{\phi}}\bra{\psi_{\phi}}\}$ is the conditional probability of the measurement outcome $x$~\cite{Demkowicz2015}. In our discussion we consider the phase $\phi$ to be defined on the interval $-\infty < \phi < \infty$ \footnote{A phase $-\infty < \phi < \infty$ is consistent with a laser phase in atomic clocks. A diffusion of the laser phase as a phase winding outside of $-\pi < \phi < \pi$ corresponds to a phase slip which must be avoided in clock operation.}.

In order to find an interferometer performing the most accurate measurement of the phase $\phi$ we cannot minimize the MSE~\eqref{eq:estimator_variance_POVM} for all values of $\phi$ simultaneously. First, the atomic interferometer is only sensitive to phase values modulo $2\pi$ as $\exp(-i\phi J_{z})$, and hence also $p(x|\phi)$ is periodic. Thus, it can not distinguish arbitrary phases. Second, an initial state and measurement working well for one phase value might be insensitive to another value. Thus we consider an estimation error minimized for a weighted range of phase values relevant for a given sensor and measurement task. In the following we will adopt a Bayesian approach where the estimation error is averaged over a prior phase distribution $\mathcal P(\phi)$. The cost function of interest is thus defined as the MSE averaged over the prior distribution, defining the Bayesian mean squared error (BMSE)
\begin{align}
    {\cal C}\equiv (\Delta \phi)^2 = \int_{-\infty}^{\infty} d \phi\,  \epsilon(\phi)\mathcal{P}(\phi).
    \label{eq:BMSQE_POVM}
\end{align}
The prior distribution $\mathcal P(\phi)$ reflects the statistical properties of the unknown phase $\phi$ hence it is, in general, sensor and task dependent.

Optimal interferometry is based on minimizing the cost function~\eqref{eq:BMSQE_POVM} over $\ket{\psi_{\rm in}}$, $\{\Pi_x\}$, and $\phi_{\rm est}(x)$ for the given prior distribution. For simplicity, we will focus on prior distribution as a normal distribution centered around zero 
\begin{equation}\label{eq:normal}
 \mathcal{P}_{ \delta \phi}(\phi)= \frac{1}{\sqrt{2\pi (\delta\phi)^2}}\exp\left[{ - \frac{\phi^2}{2(\delta\phi)^2} }\right].   
\end{equation}
This problem was addressed in~\cite{Macieszczak2014}, where the optimal quantum interferometer has been identified. Below we optimize the cost function~\eqref{eq:BMSQE_POVM} within a variational quantum algorithmic approach.

\subsection{Variational Ramsey interferometry}
\label{subsec:VariationalInterferometer}

Our goal is to find an implementation of the optimal interferometer given a restricted set of quantum gates available on an experimental platform such as neutral atoms or trapped ions. We will show that low-depth variational quantum circuits of given depth [see Fig.~\ref{fig:Fig1}(c)] are excellent approximations to optimal interferometry, and can yield significant improvements over SQL defined for uncorrelated atoms. 

In the most general form the variational interferometer, illustrated in Fig.~\ref{fig:Fig1}(b), can be defined by a generic entangling unitary operation $\mathcal{U}_{\rm En}$ preparing an entangled input state from the initial product state $\ket{\psi_0}=\ket{\downarrow}^{\otimes N}$
\begin{align}
    \ket{\psi_{\rm in}}=\mathcal{U}_{\rm En}\ket{\psi_0},
    \label{eq:Psi_in}
\end{align}
and a decoding operation $\mathcal{U}_{\rm De}$ transforming the projective measurement of a typical observable $J_z$, with eigenbasis $\ket m$,  into a generic projection
\begin{align}\label{eq:pi}
    \Pi_{x_m} \equiv  \ket{x_m}\!\bra{x_m} = \mathcal{U}_{\rm De}^{\dagger}\ket{m}\bra{m}\mathcal{U}_{\rm De}. 
\end{align}
Here we consider the subspace spanned by the states $\ket{m}$, $m\in \{-N/2, \dots, N/2\}$, which are completely symmetric under permutations of $N$ atoms, and $\ket{\psi_0}=\ket{-N/2}$. The measurement amounts to counting the difference $m$ of atoms in state $\ket{\uparrow}$ and $\ket{\downarrow}$. As shown in~\cite{Buzek1999}, this assumption can be made without loss of generality.
The basis states $\ket{m}$ are given by the eigenstates of total spin of maximum length, $j=N/2$, thus satisfying ${\bm{J}}^2\ket{m}=j(j+1)\ket{m}$ and $J_z\ket{m}=m\ket{m}$. 
As shown in~\cite{Macieszczak2014}, the optimal POVM may be restricted to the class of standard projection von Neumann measurements $\Pi_x=\ket{x}\!\bra{x}$, $\braket{x|x'}=\delta_{xx'}$. Thus the measurement of the collective spin component $J_z$ transformed by a decoder $\mathcal{U}_{\rm De}$ represents the measurement problem in full generality.

We assume that the programmable quantum sensor provides us with a set of native resource Hamiltonians $\{H_R^{(i)}\}$. The unitaries generated by these Hamiltonians determine a corresponding native set of quantum gates as variational ansatz for $\mathcal{U}_{\rm En}$ and $\mathcal{U}_{\rm De}$. A generic example is provided by global rotations $\mathcal{R}_{\mu}(\theta)=\exp(-i \theta {J}_{\mu})$ and the infinite range one-axis-twisting (OAT) interaction~\cite{Kitagawa1993} $\mathcal{T}_{\mu}(\theta) = \exp(-i \theta J^2_{\mu})$ with $\mu=x,y,z$. Such interactions have been realized on quantum simulation platforms~\cite{Leibfried2005, Appel2009, Gross2010, Riedel2010, Leroux2010, Monz2011, Lucke2011, Sewell2012, Hamley2012, Berrada2013, Barontini2015, Zhang2015, Bohnet2016, Hosten2016a, Norcia2018, Zou2018, Pogorelov2021}, and very recently also on an optical clock transition~\cite{PedrozoPeafiel2020}.  Within this  set of gates we constrain the quantum circuits to be invariant under the spin $x$-parity transformation ensuring an anti-symmetric estimator at and around $\phi=0$ (see App.~\ref{sec:spin_x_parity}). The most general circuits satisfying the $x$-parity constraint for a fixed number $n_{\rm En}$ and $n_{\rm De}$ of layers of entangling and decoding gates are
\begin{align}
	\mathcal{U}_{\text{En}}(\boldsymbol{\theta})   = \Big[&\mathcal{R}_x(\theta_{n_{\text{En}}}^{(3)})\mathcal{T}_x(\theta_{n_{\text{En}}}^{(2)})\mathcal{T}_z(\theta_{n_{\text{En}}}^{(1)}) \cdots  \nonumber\\ 
	&\mathcal{R}_x(\theta_{1}^{(3)})\mathcal{T}_x(\theta_{1}^{(2)})\mathcal{T}_z(\theta_{1}^{(1)})\Big] \mathcal{R}_y\big(\tfrac{\pi}{2}\big),
	\label{eq:Entangler}
\end{align}
and
\begin{align}
	\mathcal{U}_{\text{De}}(\boldsymbol{\vartheta})  = \mathcal{R}_x\big(\tfrac{\pi}{2}\big) \Big[&\mathcal{T}_z(\vartheta_{1}^{(1)})\mathcal{T}_x(\vartheta_{1}^{(2)})\mathcal{R}_x(\vartheta_{1}^{(3)}) \cdots\nonumber \\
	& \mathcal{T}_z(\vartheta_{n_{\text{De}}}^{(1)})\mathcal{T}_x(\vartheta_{n_{\text{De}}}^{(2)})\mathcal{R}_x(\vartheta_{n_{\text{De}}}^{(3)})\Big].
	\label{eq:Decoder}
\end{align}
Here the subscripts on the parameters indicate the layer containing the same three gates and the superscript identifies the gate within the layer. 
The complexity of the circuit is thus classified by $(n_{\rm En},n_{\rm De})$, and we have $3 (n_{\rm En}\!+\!n_{\rm De})$ (global) variational parameters in a $(n_{\rm En},n_{\rm De})$--circuit, independent of $N$. Note that here $\mathcal{U}_{\text{En}}$ and $\mathcal{U}_{\text{De}}$ commute with particle exchange. The Hilbert space dimension for dynamics in the symmetric subspace is linear in $N$, which allows us to study theoretically the scaling for large particle numbers $N$ below --- in contrast to the case of finite range interactions in Sec.~\ref{sec:FiniteRange}.  

We note that conventional Ramsey interferometry with uncorrelated atoms corresponds to the~$(0,0)$--circuit with $\mathcal{U}_{\rm En}=\mathcal{R}_y(\pi/2)$ and $\mathcal{U}_{\rm De}=\mathcal{R}_x(\pi/2)$. Here atoms are prepared initially in a product state, or coherent spin state (CSS), and remain in a product state during the evolution in interferometer followed by measurement of $J_y$. On the other hand, the interferometer with SSS as input, and GHZ interferometry emerge as $(1,0)$-- and $(2, 1)$--circuits, respectively. 

In the presented entangler-decoder framework the performance of the interferometer is described, similar to Eq.~\eqref{eq:estimator_variance_POVM}, by the MSE
\begin{equation}
    \epsilon(\phi) =\sum_{m} \big[\phi-\phi_{\rm est}(m)\big]^2 p_{\boldsymbol{\theta},\boldsymbol{\vartheta}}(m|\phi),
    \label{eq:estimator_variance}
\end{equation}
where the conditional probability is
\begin{align}
    p_{\boldsymbol{\theta},\boldsymbol{\vartheta}}(m|\phi) = |\bra{m}\mathcal{U}_{\rm De}(\boldsymbol{\vartheta}) e^{-i\phi J_z} \mathcal{U}_{\rm En}(\boldsymbol{\theta})\ket{\psi_{0}}|^2.
    \label{eq:MeasurementProbabilityDistribution}
\end{align}
Therefore, the optimal interferometer found within the restricted set of available operations is described by the minimum of the BMSE
\begin{align}
     (\Delta \phi )^2 = \min_{\boldsymbol{\theta},\boldsymbol{\vartheta},a} \int_{-\infty}^{\infty} d \phi \,  \sum_m&(\phi-am)^2 \notag \\ &\times p_{\boldsymbol{\theta},\boldsymbol{\vartheta}}(m|\phi)\mathcal{P}_{\delta \phi}(\phi).
    \label{eq:BMSQE}
\end{align}
To be specific, we assume for the prior  a normal distribution $\mathcal{P}_{\delta \phi}(\phi)$  with standard deviation $\delta \phi$ [see Eq.~\eqref{eq:normal}]. In addition, (\ref{eq:BMSQE}) assumes a linear estimator $\phi_{\rm est}(m) = am$ which is close to optimal, as shown below. We note that it is possible to use the optimal Bayesian estimator, which however is  computationally demanding. We describe the corresponding iterative procedure in App.~\ref{app:covarinat_measurement} for the case of a phase operator as observable.

\begin{figure}[t]
    \centering
    \includegraphics[]{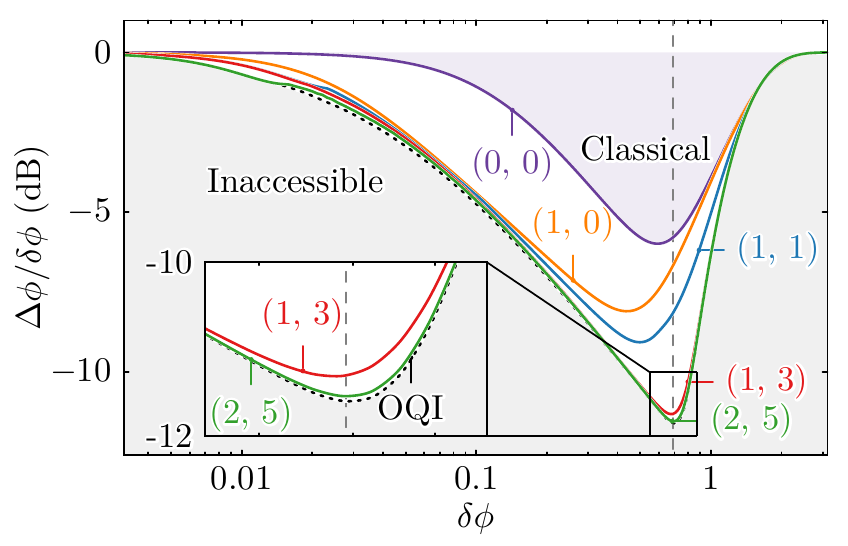}
    \caption{Performance of the variationally enhanced interferometer with $N=64$ particles. Performance is shown in terms of the posterior phase distribution width relative to the prior width, $\Delta\phi/\delta\phi$, for a given prior, that is, for a given dynamic range of the interferometer. 
    Colored lines show the performance of variationally optimized circuits for the depth $(n_{\rm En}, n_{\rm De})$ of entangling and decoding layers as indicated. The number of variational parameters is given by $3(n_{\rm En}+ n_{\rm De})$. 
    The performance of the optimal quantum interferometer (OQI)~\cite{Macieszczak2014} is indicated by the dotted line. The shaded areas indicate the classically accessible~(purple) and the quantum mechanically forbidden~(gray) regions (for $N=64$). Related results applied to atomic clocks are shown in Fig.~\ref{fig:AllanDeviation_hT}}
    \label{fig:results}
\end{figure} 

\begin{figure*}[t!]
    \centering
    \includegraphics[]{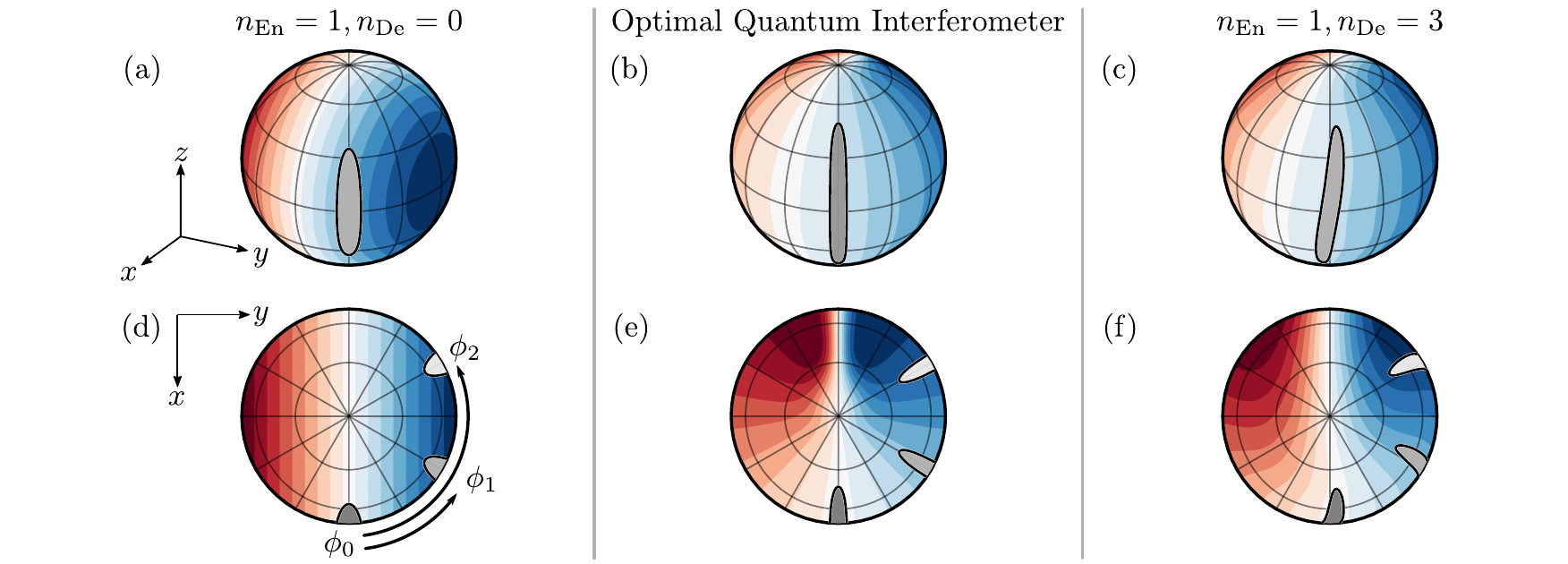}\\
    \includegraphics[]{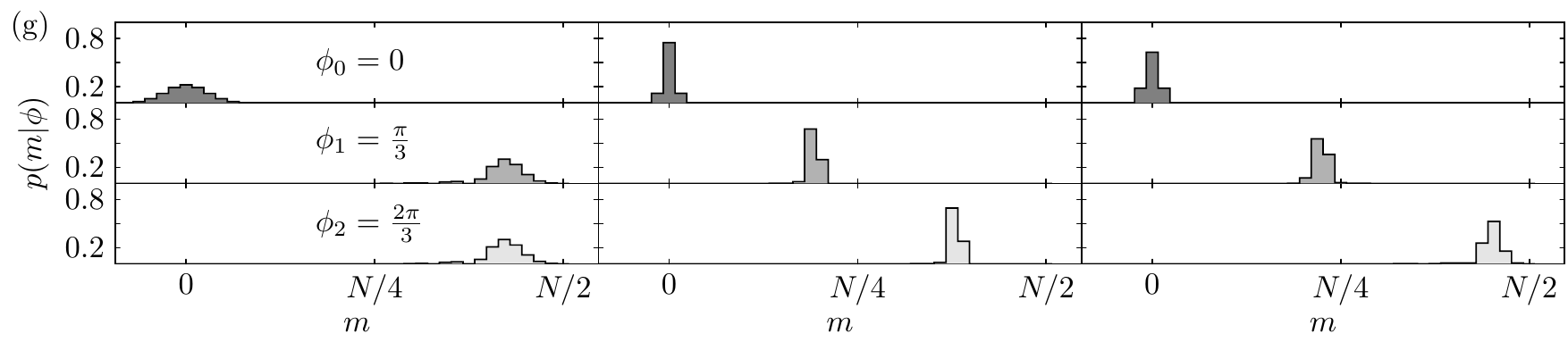}
    \caption{Visualization of quantum states $\ket{\psi_{\phi}}=\exp(-i\phi J_{z})\ket{\psi_{\rm in}}$, and quantum measurement operators as Wigner distributions on the generalized Bloch sphere for $N=64$ and $\delta \phi\approx 0.7$. The first  (a,d), second  (b,e), and third column (c,f) correspond to $(n_{\rm En},n_{\rm De})=(1,0)$ (squeezed input state, and  $J_y$ measurement operator), the optimal quantum interferometer, and to a $(1,3)$ quantum circuit, respectively. Measurement operators are visualized as colored contours on the Bloch sphere corresponding to different measurement outcomes. The corresponding optimized (optimal) states $\ket{\psi_{\phi}}$ are shown at various angles $\phi$ as gray shaded areas. (a,b,c) three dimensional view of the generalized Bloch sphere with a state rotated to $\phi= \pi/3$. (d, e, f) Top view of the Bloch sphere with the state rotated to angles $\phi=\ 0,\ \pi/3,\ 2\pi/3$. (g) Measurement probability $p(m|\phi)$ [see Eq.~\eqref{eq:MeasurementProbabilityDistribution}] corresponding  to the overlap between the contours of the measurement distribution and the respective state distribution, displayed in the same column. The three rows correspond to the above three angles $\phi$. Note that for  the $J_y$ measurement the distributions at angles $\pi/3$ and $2\pi/3$ are indistinguishable in measurement statistics. In contrast, for the OQI and the $(1,3)$ quantum circuit these angles are well resolved.}
    \label{fig:Wigner_distributions}
\end{figure*}

\subsection{Results of optimization}\label{sec:ResultsOptimization}

Results of interferometer optimizations \footnote{For the optimization results presented throughout this manuscript, we calculate exact gradients and use a combination of sequential quadratic programming and the Nelder-Mead method, as implemented in \textsc{Matlab}, to determine optimal solutions.} are shown in Fig.~\ref{fig:results} for $N=64$ atoms. The figure plots the ratio $\Delta\phi/\delta \phi$ of the root BMSE $\Delta \phi$ relative to the normal prior width $\delta \phi$. The more information  we gain about the parameter $\phi$ in a single measurement, the smaller the value of this ratio.

The black dotted line shows the result of the unrestricted minimization of the cost function~\eqref{eq:BMSQE_POVM} with normal prior~\cite{Macieszczak2014}, which we refer to as optimal quantum interferometer (OQI). It defines the region (shaded area) inaccessible to any $N$-particle quantum interferometer. The purple line represents performance of the conventional Ramsey interferometer with CSS as input and a linear estimator, given by the $(0,0)$--circuit. Thus, the shaded area above the purple line roughly defines the classically achievable performance.

The performance of the entanglement enhanced interferometer is shown with colored lines. The orange curve represents a $(1,0)$--circuit corresponding to a squeezed spin state (SSS) interferometer~\cite{Wineland1992, Kitagawa1993}, employing the OAT interaction to generate an entangled initial state with suppressed fluctuations along the axis of the effective $J_y$ measurement. The minimum of the orange line is located at smaller $\delta \phi$ values when compared to the minimum of the purple line corresponding to the SQL. This manifests the fact that SSS input state increases the sensitivity of the phase measurement at expense of dynamic range~\cite{Andre2004,He2011, Braverman2018}. By adding a single layer of a decoding circuit we obtain the blue curve corresponding to the $(1,1)$--interferometer with a slightly enhanced sensitivity and dynamic range. The red and green lines correspond to $(1,3)$-- and $(2,5)$--circuits, respectively, and show striking improvement in  sensitivity, providing an excellent approximation for the optimal interferometer (black dotted line). Remarkably, the minima of the red, green, and black curves are located at a wider dynamic range $\delta \phi$ than that of the CSS interferometer. Hence the optimal entangled initial state and the effective nonlocal observable allows us to achieve both \emph{a higher phase sensitivity} and \emph{a wider dynamic range}. 

\begin{figure}[t]
    \centering
    \includegraphics[width=\columnwidth]{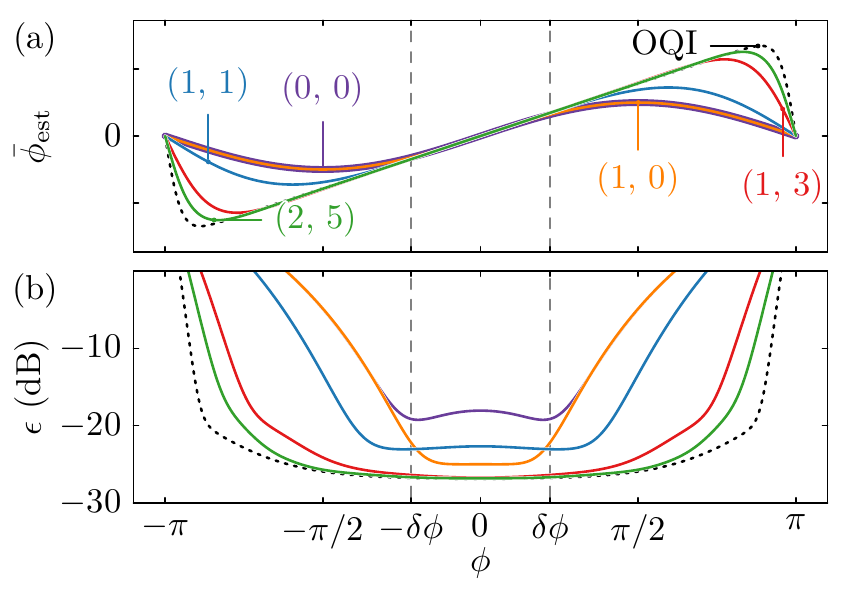}
    \caption{(a) Phase $\phi$ dependence of the estimator expectation value [Eq.~\eqref{eq:EstimatorExpectationValue}] of the optimized $N$ = 64 particle interferometer at different circuit depths $(n_{\rm En}, n_{\rm De})$ with $3(n_{\rm En}+ n_{\rm De})$ variational parameters,  in comparison to the optimal quantum interferometer (OQI).  The optimization is performed for the prior distribution width $\delta \phi \approx 0.7$, indicated by the vertical lines. (b) Mean squared error [Eq.~\eqref{eq:estimator_variance}] corresponding to the estimator expectation values curves above.}
    \label{fig:dynamicRange}
\end{figure}

To gain understanding of the physical meaning of the measurements and initial states emerging from the numerical optimization, we show their Wigner functions in Fig.~\ref{fig:Wigner_distributions}. A formal definition of the Wigner distribution is provided in App.~\ref{sec:WignerDistribution}. The three columns correspond, in consecutive order, to the $(1,0)$--circuit (SSS interferometer), the optimal quantum interferometer of~\cite{Macieszczak2014}, and the $(1,3)$--circuit. The chosen prior width $\delta \phi\approx 0.7$ is indicated in Fig.~\ref{fig:results} by the vertical dashed line. The first row of panels~\ref{fig:Wigner_distributions}(a-c) shows 3D views of the generalized Bloch sphere with Wigner functions of the measurement operators shown in shades of red and blue for $J_y$, the optimal observable, and $\smash{\mathcal{U}_{\rm De}^{\dagger}J_z\mathcal{U}_{\rm De}}$ operators, respectively. A contour of constant color corresponds roughly to a certain measurement outcome which is obtained with the probability given by the overlap of the contour with the Wigner function of a quantum state. The states are shown in \ref{fig:Wigner_distributions}(a-f) with the gray outlined areas.

Panel~\ref{fig:Wigner_distributions}(a) shows clearly the non-optimality of the SSS interferometer with a measurement of the spin projection $J_y$. Optimization of the SSS results in a  moderate level of squeezing (gray ellipse squeezed along the $y$-axis). More squeezing would produce stronger anti-squeezing along $z$-axis leading to overlap with more contours of the $J_y$ Wigner function, thus increasing the variance of the measurement results for nonzero $\phi$~\cite{Andre2004, Braverman2018}. Another limitation of the SSS interferometer, illustrated in panels~(d) and (g), is the reduced dynamic range in the interval $-\pi/2$ and $\pi/2$. Panels (d) and (g) show that states rotated by the phase angle $\phi=2\pi/3>\pi/2$ have the same measurement statistics as states rotated by $\phi=\pi/3$. Thus, phases outside the $[-\pi/2,\pi/2]$ interval can not be reliably estimated.

The optimal quantum interferometer is explained in the central column of Fig.~\ref{fig:Wigner_distributions}. Here panel~(b) shows that the initial state is squeezed significantly stronger than in the SSS interferometer. This is possible because the corresponding optimal measurement is very similar to the phase operator of Pegg and Barnett~\cite{Pegg_1988}, which has eigenstates with well defined phases (see Sec.~\ref{subsec:covariant_measurement_comparison} below for detailed comparison). One can see that the color contours of the optimal measurement Wigner function in panels (b) and (e) are aligned with the meridians and thus overlap favorably with the strongly squeezed initial state rotated by a wide range of phase angles $\phi$. Strikingly, the OQI can effectively use the full $2\pi$ dynamic range, as illustrated in panels (e) and (g).

Finally, the $(1,3)$--interferometer, presented in the third column of Fig.~\ref{fig:Wigner_distributions}, exhibits properties similar to the OQI. Interestingly, the initial state in this case is not a conventional squeezed state, as shown in panel (c), but a slightly twisted one. This, however, does not impair the performance of the interferometer as the effective measurement is also twisted such that it matches the initial state rotated by a wide range of phase angles. This peculiarity is a consequence of the restricted gate set available for the variational optimization in a realistic system. It is remarkable that the low depth $(1,3)$--circuit already provides an excellent approximation for the OQI.

The extended dynamic range of the variationally optimized interferometer is explored in Fig.~\ref{fig:dynamicRange}. Panels (a) and (b) show, respectively, the estimator expectation value 
\begin{align}
\bar \phi_{\rm est}\equiv\sum_{m}\phi_{\rm est}(m)p(m|\phi)
\label{eq:EstimatorExpectationValue}
\end{align}
and the estimator mean squared error~\eqref{eq:estimator_variance} as functions of the actual phase $\phi$ for an interferometer optimized for prior width of $\delta \phi\approx0.7$ (indicated with vertical dashed lines).

The estimator expectation value of the $(0,0)$-- and $(1,0)$--circuit (CSS and SSS interferometer) is given by a sine function [purple and orange line in panel~(a)], thus, it can unambiguously map the estimated phase to the actual phase in the range between $-\pi/2$ and $\pi/2$. However, the useful dynamic range of the interferometer is even narrower as shown by the estimator error in panel~(b). The estimator error of the SSS state is suppressed below the CSS benchmark line only for phases between, roughly, $-\pi/4$ and $\pi/4$. The $(1,1)$--interferometer [blue line in (a) and (b)] starts to exploit the entangled measurement and achieves a bit wider linear regime of $\bar \phi_{\rm est}$ in (a) and a wider region of suppressed estimator error in (b). Although the minimum error of $(1,1)$--circuit is larger than that of $(1,0)$--circuit, it still has superior overall sensitivity as phases in the tails of the prior distribution are better resolved. 

Finally, more complex decoding operations employed by the $(1,3)$-- and $(2,5)$--circuit (red and green lines) allow to approach the performance of the optimal interferometer (black dotted lines). The linear regime of $\bar \phi_{\rm est}$ extends almost to the full $2\pi$ range, and the estimator error is well suppressed for phases deeply within the tails of the prior.

\subsection{Comparison between variational and phase operator based interferometers}\label{subsec:covariant_measurement_comparison}

\begin{figure}[t] 
   \centering
   \includegraphics[width=\columnwidth]{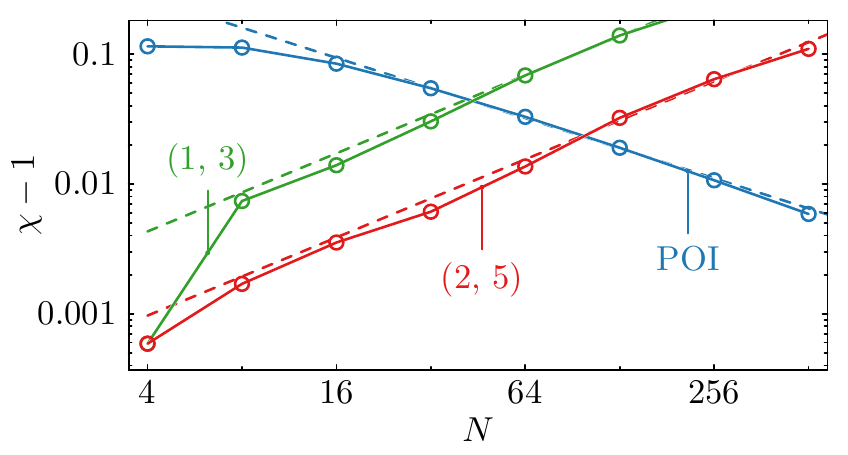} 
   \caption{Relative performance of the covariant, phase operator based interferometer (POI) (blue) and the variational $(1,3)$-- and $(2,5)$--interferometers (green and red points, respectively) with respect to the OQI for a given system size $N$. The $\chi$ ratio is defined in Eq.~\eqref{eq:chi}, OQI corresponds to $\chi=1$. The dashed lines represent empirical scalings, green and red one scale as~$\sim N$, and blue is $\sim N^{-0.77}$.}
   \label{fig:covariant_measurement}
\end{figure}

From a theory perspective it is interesting to compare the performance of the variationally optimized interferometer and the  interferometer based on covariant measurement~\cite{Holevo1982,Buzek1999}. Here covariant measurements represent the class of measurements optimal for phase estimation with no \emph{a priori} knowledge and phase-shift symmetry, i.e.~assuming a prior distribution $\mathcal{P}(\phi)=(2\pi)^{-1}$ and a $2\pi$-periodic cost function, as opposed to the MSE~\eqref{eq:estimator_variance_POVM}.

In the case of clocks and magnetometry, the free evolution encoding the phase $\phi$ is the collective spin rotation, $e^{-i\phi J_z}$. The corresponding covariant measurement optimal for estimation of the rotation angle $\phi$ can be represented by the von Neumann measurement~\cite{Derka1998} with phase operator $\hat\Phi$~\cite{Pegg_1988}, which we define in App.~\ref{app:covarinat_measurement}.

In order to evaluate the performance of the phase operator based interferometer (POI), we minimize the cost function~\eqref{eq:BMSQE_POVM} for $\hat\Phi$ as the observable and the normal prior $\mathcal{P}_{\delta\phi}(\phi)$.  To this end, we use the optimal Bayesian estimator known as the \emph{Minimum Mean Squared Error} (MMSE) estimator~\cite{Demkowicz2015} and find the corresponding optimal initial state $\ket{\psi_{\hat\Phi}}$ (see App.~\ref{app:covarinat_measurement} for details). This results in the optimal posterior width $\Delta\phi_{\rm POI}$ as discussed in Sec.~\ref{sec:ResultsOptimization} for the variationally optimized interferometer.

To compare different interferometers we consider their performance at the optimal prior width with respect to the OQI performance and define the ratio:
\begin{equation}
\label{eq:chi}
\chi = \frac{\min_{\delta\phi}(\Delta\phi/\delta\phi)}{\min_{\delta\phi}(\Delta\phi_{\rm OQI}/\delta\phi)}.
\end{equation}
The $\chi$ value corresponds to the ratio of minima of an interferometer and the OQI curves in Fig.~\ref{fig:results}. The OQI corresponds to $\chi=1$.

Figure~\ref{fig:covariant_measurement} shows the $\chi-1$ value for variationally optimized and $\hat\Phi$ based interferometers for various system sizes up to  $N=512$. The figure highlights sub-optimality of the POI (blue points) for the task of phase estimation with \emph{non-periodic} cost function, as is relevant for frequency estimation in, e.g., optical clocks. For small systems, $N \lesssim 16$, the POI is up to $\sim 10\%$ less efficient than the OQI and the variational $(1,3)$-- and $(2,5)$--interferometers (green and red points, respectively). $(1,3)$--circuit outperforms POI for systems of up to $N\sim40$ atoms, whereas $(2,5)$--circuit is better for up to $N\sim 100$ atoms. In the limit of large number of atoms, $N\ggg 1$, the POI approaches the OQI performance. Empirical fitting indicates convergence rate $\chi_{\rm POI}-1\sim N^{-0.77}$, as $N$ increases. On the other hand, the variationally optimized interferometers diverge from OQI linearly with $N$.

\subsection{Variational Optimization in Presence of Imperfections and Noise}\label{subsec:dephasing}

Variational optimization can be extended to include imperfections and decoherence. This optimization can also be carried out on the physical quantum sensor. This is particularly beneficial when the experimental characterization of imperfections and noise is incomplete. 

There are various sources of imperfections and decoherence, which are relevant in our context. First, there are control errors in implementing variational quantum gates. These include offsets of control parameters and Hamiltonian design errors. The latter are deviations of the physically realized  vs.~the ideal Hamiltonian, e.g.~in the implementation of one-axis twisting interaction. However, if these (unknown) control or design errors are static, i.e.~do not fluctuate between experimental runs, a variational algorithm performed on the device will still optimize, and thus compensate in the best possible way for these errors in $\mathcal{U}_{\text{En}}$ and $\mathcal{U}_{\text{De}}$, i.e.~find the best gate decomposition for given building blocks. In addition, there will be decoherence due to fluctuations of control parameters, or coupling to an environment as in spontaneous emission or dephasing. 

To incorporate the latter we need to extend the formalism to density matrices instead of the previously discussed pure states. 
Below we illustrate this by an optimization of the Ramsey interferometer in the presence of single atom dephasing noise during the Ramsey interrogation time $T$, as one example of experimentally relevant decoherence. Local dephasing noise is described by the Lindbladian
$\mathcal{L}\circ\rho = (1/4)\sum_{j=1}^{N}\left( \sigma^z_j \rho \,\sigma^z_j- \rho \right)$.
Thus the density matrix after the Ramsey interrogation time, 
\begin{align}
	\rho_{\bm{\theta}}^{\phi, \gamma  T} = e^{-i \phi J_z}\left( e^{ \gamma T\mathcal{L}\circ}\rho_{\bm{\theta}}\right)e^{i \phi J_z},
\end{align}
can be expressed in terms of the dimensionless phase $\phi$ accumulated during the Ramsey interrogation time $T$ and the effective exposure to the dephasing noise $\gamma  T$ with dephasing rate $\gamma$. Here $\rho_{\bm{\theta}} = \mathcal{U}_{\rm En}(\bm{\theta})\ket{\psi_{0}}\bra{\psi_{0}}\mathcal{U}^{\dagger}_{\rm En}(\bm{\theta})$, where we used that the dephasing Lindbladian and the free evolution of the clock supercommute. The particle permutation symmetry of the Lindbladian enables us to simulate systems at a cubic cost in $N$~\cite{Chase2008, Shammah2018}. 
The conditional probability, required to determine the BMSE in Eq.~\eqref{eq:BMSQE} therefore reads
\begin{align}
     p_{\boldsymbol{\theta},\boldsymbol{\vartheta}}(m|\phi, \gamma  T) = \bra{m}\mathcal{U}_{\rm De}(\boldsymbol{\vartheta}) \rho_{\bm{\theta}}^{\phi, \gamma  T} \mathcal{U}^{\dagger}_{\rm De}(\boldsymbol{\vartheta})\ket{m}.
\end{align}

\begin{figure}[t]
    \centering
    \includegraphics[width=\columnwidth]{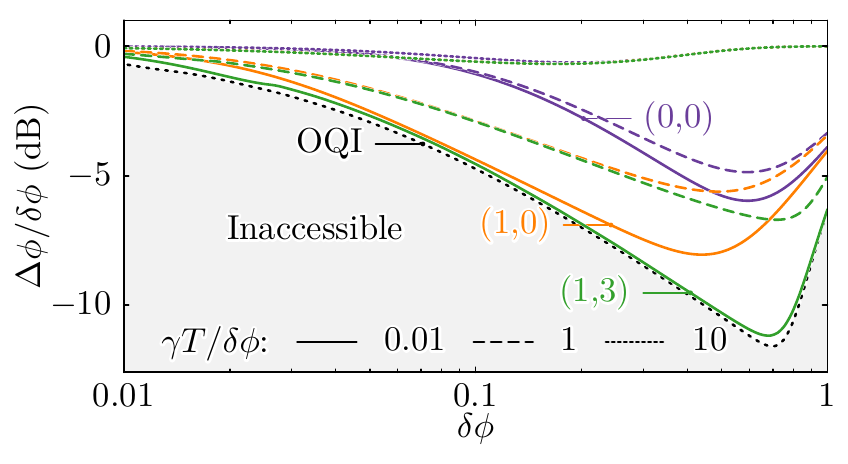}
    \caption{Accuracy of an optimized $N=64$ particle interferometer in the presence of single particle dephasing noise exposures $\gamma  T/\delta \phi$ relative to the prior distribution width,  indicated by different line styles. The results are displayed for different circuit complexities $(n_{\rm En},n_{\rm De})$ with $3(n_{\rm En}+ n_{\rm De})$ variational parameters. For comparison the accuracy of the noise free optimal quantum interferometer (OQI) is indicated by the black dotted line.  }
    \label{fig:posterior_prior_dephasing}
\end{figure}

Figure~\ref{fig:posterior_prior_dephasing} shows that the optimized $\Delta \phi/\delta \phi$ increases as the noise increases, as expected. For a small  $\gamma  T/\delta \phi = 0.01$ the variational $(1, 3)$--interferometer is close to optimal without noise. Remarkably for all ratios $\gamma  T/ \delta \phi\lesssim 1$, the minimum of the $(1, 3)$ interferometer remains well below the uncorrelated $(0,0)$-- and the SSS $(1,0)$--interferometers. This ordering of the respective global minimum is independent of $N$,
whereas for $\gamma  T/\delta \phi = 10$ none of the entangling sequences improve significantly compared to SQL~\cite{Huelga1997}. 

\subsection{Towards the Heisenberg limit}\label{subsec:HL}

The variationally optimized interferometer with low-depth quantum circuits
found within the Bayesian framework quickly approaches the accuracy
of the optimal Ramsey interferometer.
We will now discuss our results from the perspective of reaching the
Heisenberg limit (HL).

The HL is a lower bound on the accuracy of an interferometer imposed
by quantum mechanics. For an $N$-atom interferometer the HL and SQL
are traditionally written as 
\begin{align}
  \Delta\phi_{{\rm HL}}^2\ge\frac{1}{N^2},\quad\Delta\phi^2_{{\rm SQL}}\ge \frac{1}{N},  
\end{align}
which must be understood in context of the quantum Fisher information~\cite{Braunstein1994, Braunstein1996, Paris2009}
and quantum Cram\'er-Rao bound~\cite{Helstrom1976, Holevo1982} (implying $\delta\phi\rightarrow 0$). In contrast, in the present work we
have adopted a Bayesian approach,
which includes optimizing for a \emph{finite dynamic range} $\delta \phi$. To evaluate the performance of our quantum variational results for a given circuit depth in comparison with HL, we will adopt below van Trees inequality~\cite{Trees1968, Gill1995} as a bound for the BMSE.

In brief, for any given conditional probability distribution $p(m|\phi)$ the Cram\'er-Rao inequality 
\begin{equation}
    V(\phi) \geq \frac{1}{F_{\phi}}
    \label{eq:CR_bound}
\end{equation}
provides a bound on the variance of an unbiased ($\bar{\phi}_{{\rm est}}=\phi$) estimator \mbox{$
V(\phi) \equiv \sum_{m}[\phi_{{\rm est}}(m)-\bar{\phi}_{{\rm est}}]^{2}p(m|\phi)
$}
based on the Fisher information
\begin{equation}
  F_{\phi}=\sum_{m}\big[\partial_{\phi}\log p(m|\phi)\big]^{2}p(m|\phi).
  \label{eq:Fisher_info}
\end{equation}
For pure states, i.e.~in the absence of decoherence, $\smash{F_{\phi}\leq N^{2}}$ in correspondence to the HL above. We emphasize that the Cram\'er-Rao inequality seeks to identify optimal unbiased estimators, which can in general
be achieved only \emph{locally} in $\phi$, i.e.~in a
small neighborhood of a given phase, and not for a finite dynamic range as is the goal in our Bayesian approach.

\begin{figure}[t]
    \centering
    \includegraphics[width=\columnwidth]{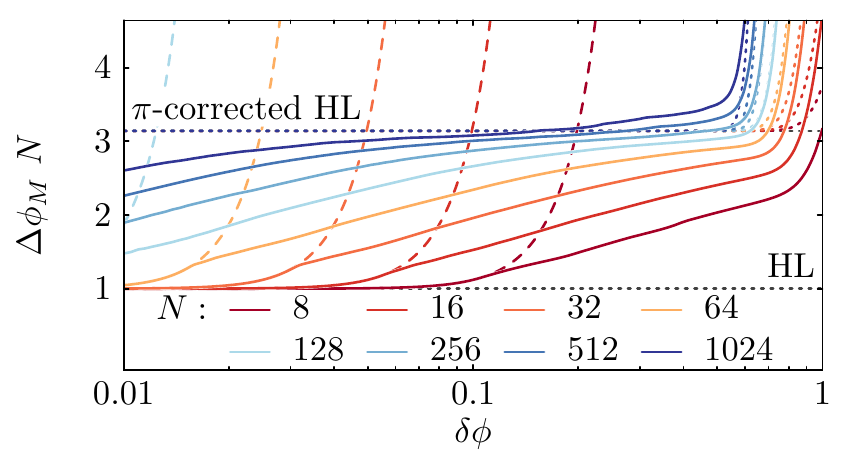}
    \caption{Plot of $\Delta \phi_M\, N$, i.e.~the standard deviation of an effective measurement rescaled by the ensemble size $N$, vs.~prior width $\delta \phi$. 
    Solid lines show results for the optimized interferometer with circuit depth $(n_{\rm En},n_{\rm De})=(2, 5)$ in comparison to the analytic expression describing a GHZ-state interferometer [Eq.~\eqref{eq:GHZ_effective_mes_variance}] shown with dashed lines and the \emph{$\pi$-corrected Heisenberg limit} including phase slips [Eq.~\eqref{eq:piHL_effective_mes_variance}] shown with dotted lines. The \emph{Heisenberg limit} and the \emph{$\pi$-corrected Heisenberg limit} are indicated with dotted horizontal lines.}
    \label{fig:Heisenberg_limit}
\end{figure}

\begin{figure*}[t]
    \centering
    \includegraphics[]{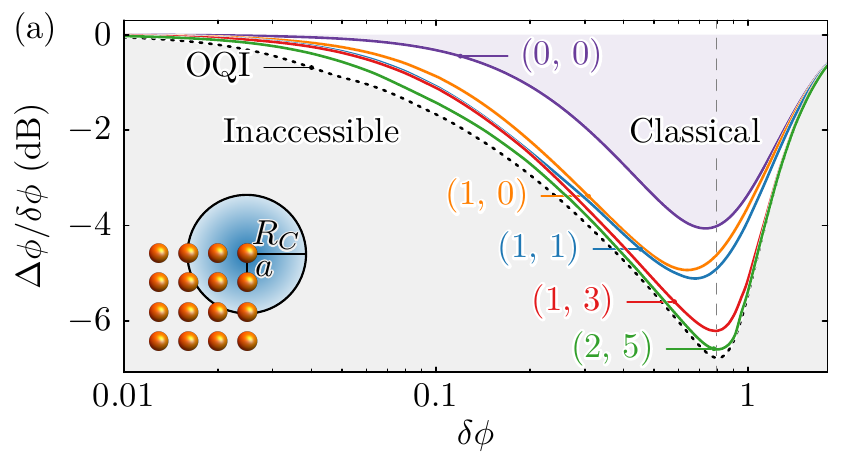}
    \includegraphics[]{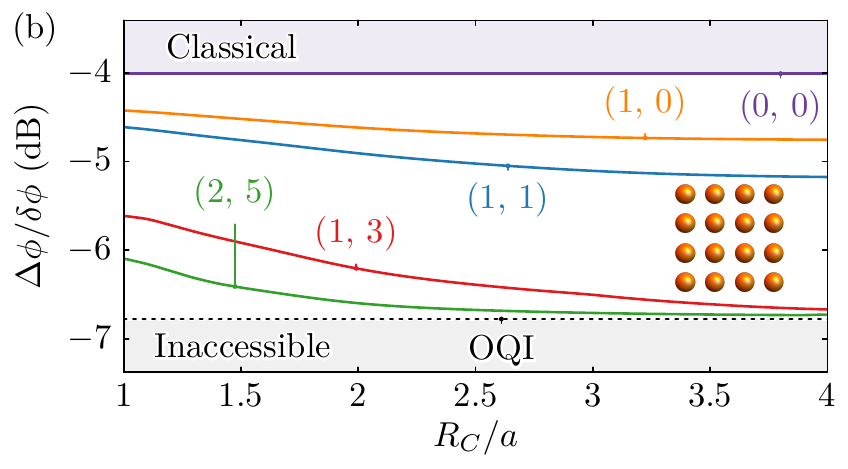}
    \caption{Performance of the variationally optimized $N=16$ Ramsey interferometer on a $4\times 4$ lattice interacting via finite range Rydberg dressing interactions \eqref{eq:Vfiniterange} with interaction radii $R_C$ in units of the array spacing $a$. Colored lines show the reduction of the posterior phase distribution width $\Delta\phi$ relative to the prior distribution width $\delta\phi$ for variationally optimized circuits complexity $(n_{\rm En}, n_{\rm De})$ with $3(n_{\rm En}+ n_{\rm De})$ variational parameters. The performance of the optimal quantum interferometer (OQI)~\cite{Macieszczak2014} is indicated by dotted lines. The shaded areas indicate the classically accessible~(purple) and the quantum mechanically inaccessible~(gray) regions. (a) Prior width  dependence of the optimized solution at $R_{C}/a=2$. (b) Interaction range dependence of the optimal solution at a prior distribution width $\delta\phi\approx0.8$ indicated by the vertical dashed line in (a).}
    \label{fig:Dressing}
\end{figure*}

In the Bayesian framework, a bound on the BMSE 
is imposed by van Trees' inequality, 
\begin{align}
\big(\Delta\phi\big)^{2} \geq  \frac{1}{ \overline{F}_{\phi} + \mathcal{I}}.
\end{align}
Here, the first term in the denominator is the Fisher information~\eqref{eq:Fisher_info} averaged over the prior distribution, $\overline{F}_{\phi}=\int d\phi\mathcal{P}(\phi) F_{\phi}$. The second term is the Fisher information
of the prior distribution, $\mathcal{I}=\int d\phi\mathcal{P}(\phi)\big[\partial_{\phi}\log\mathcal{P}(\phi)\big]^{2}$, representing the prior knowledge.
To isolate the measurement contribution from the prior knowledge,
we define an effective measurement variance $(\Delta\phi_{M})^{2}$ via
\begin{equation}
\frac{1}{(\Delta\phi_{M})^{2}}\equiv \frac{1}{(\Delta\phi)^{2}}-\mathcal{I},
\label{eq:effective_measurement_variance}
\end{equation}
and obtain
\begin{equation}
(\Delta\phi_{M})^{2}\geq\frac{1}{\overline{F}_{\phi}}\geq\frac{1}{N^{2}},
\label{eq:BCR_bound}
\end{equation}
reminiscent  of the Cram\'er-Rao inequality~\eqref{eq:CR_bound}.  In 
case of a normal prior distribution~\eqref{eq:normal} we have \mbox{$\mathcal{I}=(\delta\phi)^{-2}$}, and the effective measurement variance~\eqref{eq:effective_measurement_variance} reads \mbox{$(\Delta\phi_{M})^{-2}\equiv (\Delta\phi)^{-2}-(\delta\phi)^{-2}$}.

In Fig.~\ref{fig:Heisenberg_limit} we plot $(\Delta\phi_{M}) \times N$, the measurement error scaled to the atom number,  for the $(2,5)$--variational interferometer
(solid lines) as a function of the prior width $\delta \phi$ for a range of atom numbers $N$. In addition, we indicate the HL and the $\pi$-corrected HL (see below) as dotted lines and show results for a GHZ interferometer with
spin $x$-parity measurement~\cite{Bollinger1996} (dashed lines).
In the case of the GHZ interferometer with a normal prior we have
\begin{equation}
(\Delta\phi_{M}^{{\rm GHZ}})^{2}=\frac{e^{(N\delta\phi)^{2}}}{N^{2}}-(\delta\phi)^{2}
\label{eq:GHZ_effective_mes_variance}
\end{equation}
showing that
the GHZ interferometer attains the HL uncertainty $\Delta\phi_{M}\rightarrow1/N$
for a given prior width $\delta\phi$ only for atom numbers $N\lesssim1/\delta\phi$.
This fact is illustrated in Fig.~\ref{fig:Heisenberg_limit} by the
dashed lines which diverge from the HL for smaller and smaller $\delta\phi$
as $N$ grows. In contrast, the variational interferometer (solid
lines) is of the order of the $\pi$-corrected HL~\cite{Buzek1999,Berry2000,Jarzyna2015,Gorecki2020}, $\Delta\phi_{M}\rightarrow\pi/N$,
for a \textit{wide range} of prior widths $\delta\phi$ as $N$ increases. 

Intuitively, the emergence of $\pi$-corrected HL can be understood as follows. The optimal $N$ atom quantum interferometer can be described as a von Neumann measurement in the particle permutation symmetric subspace~\cite{Buzek1999,Macieszczak2014}. Thus, there are $N+1$ possible measurement outcomes to distinguish at most $N+1$ phase values in the interval $[-\pi,\pi]$. The corresponding estimation error for evenly spread estimates reads $\Delta\phi\sim (1/2)\,2\pi/(N+1)\to\pi/N$.

For large $\delta \phi$ the solid lines in Fig.~\ref{fig:Heisenberg_limit} exhibit strong deviations from the asymptotic  $\pi$-corrected HL behavior. The cusps are explained by phase slips outside the interval $[-\pi, \pi]$ which lead to a squared estimation error of $4 \pi ^ 2$. For a normal prior distribution, the performance of an interferometer limited by the $\pi$-corrected HL including the phase slips is given by 

\begin{equation}
(\Delta\phi_{M}^{{\pi\rm HL}})^{2}=\frac{\pi ^ 2}{N ^ 2} + 4 \pi ^2 \Big(1-\mathrm{erf}\frac{\pi}{\sqrt{2} \delta \phi}\Big).
\label{eq:piHL_effective_mes_variance}
\end{equation}
Results of this section are obtained in absence of decoherence.

\subsection{Finite range interactions}
\label{sec:FiniteRange}
Our previous discussion assumed infinite range interactions as entangling quantum resource, while e.g.~neutral atoms stored in tweezer arrays feature finite range interactions.  The variational optimization of the BMSE can be directly generalized to finite range interactions, which we  illustrate by optimizing a sensor based on Rydberg dressing resources~\cite{Henkel2010, Pupillo2010} $\mathcal{D}_{\mu} (\theta) = \exp [{-i \theta (H^{D}_{\mu}/V_0)}]$, as is realized in alkaline earth tweezer clocks~\cite{Covey2019, Wilson2019, Young2020}. The effective interaction Hamiltonian we use for the optimization reads 
\begin{align}\label{eq:Vfiniterange}
    H^{D}_{\mu} = \sum_{k,l=1}^{N}\frac{V_0 R_{C}^6/4}{|\bm{r}_k-\bm{r}_l|^{6}+R_C^6}{\sigma}^{\mu}_k{\sigma}^{\mu}_l,\quad   (\mu=x,y,z)
\end{align}
where $\bm{r}_k$ represents the position of particle $k$. The interaction strength at short distances $V_0$ and interaction radius $R_{C}$ depend on the Rydberg level and the dressing laser used to let the particles interact~\cite{Gil2014}. 

Ref.~\cite{Kaubruegger2019} presented a study of variationally optimized spin-squeezed input states, and we refer to this work for the elementary gates we employ as building blocks for variationally optimizing entangling and decoding operations. In analogy to Eqs.~\eqref{eq:Entangler} and \eqref{eq:Decoder}, we write
the entangler and decoder, effectively replacing the ${\cal T}_{x,z}$ by ${\cal D}_{x,z}$. In a similar way we can rewrite Eq.~\eqref{eq:MeasurementProbabilityDistribution} to account for dynamics in full $2^N$-dimensional Hilbert space.

Figure~\ref{fig:Dressing}(a) shows the optimized $\Delta \phi/\delta \phi$ for a $4\times4$ square array for  $R_C=a$ with $a$ the lattice constant. We find variational solutions approximating the OQI, similarly to the OAT interactions in Fig.~\ref{fig:results}. In contrast to the infinite range OAT interaction we are not able to exactly reproduce the optimal GHZ-state interferometer at $\delta \phi < 1 / N$. Nonetheless, at any prior distribution width significant improvement beyond the uncorrelated interferometer is achieved and in particular around global minimum of the optimal interferometer (vertical dashed line), the decoder-enhanced circuits clearly surpass sensitivity of entangled input states only. 

In Fig.~\ref{fig:Dressing}(b) we further study the dependence on the scaled interaction radius $R_C/a$ for a fixed prior distribution width $\delta \phi$ corresponding to the minima of variational and optimal interferometer curves in Fig.~\ref{fig:Dressing}(a) (vertical dashed line). We see that even in the limit of an effective nearest neighbor interaction $R_C=a$ a clear improvement beyond  the classical sensitivity limit is possible. As the interaction radius increases, the root BMSE of the variationally optimized interferometer decreases, ultimately reproducing the results of infinite-range interactions in the limit $R_C/a\rightarrow \infty$.

Theoretical treatment of the variational interferometry with finite range interactions involves solution of a quantum many-body problem. This, in general, is an exponentially hard problem representing the regime where variational optimization on the quantum sensor as a physical device provides a (relevant) quantum advantage, beyond the capabilities of classical computation.

\section{Application to Atomic Clocks}\label{sec:clock}

Atomic clocks realized with neutral atoms in optical trap arrays or trapped ions provide us with natural entanglement resources to implement variationally optimized Ramsey interferometry. Below we provide a study of a variationally optimized clock assuming as quantum resources global spin rotations and OAT, as realized, for example, with trapped ions as M{\o}lmer-S{\o}rensen gate, or in cavity setups with neutral atoms. This discussion is readily extended to other platforms and resources. 

Optical atomic clocks operate by locking the fluctuating laser frequency $\omega_L (t)$ to an atomic transition frequency $\omega_A $~\cite{Ludlow2015}. To this end, an atomic interferometer is used to repeatedly measure the phase $\phi_k=\int_{t_k}^{t_k+ T}dt[\omega_L (t)-\omega_A ]$ accumulated during interrogation time $T $ at the $k$-th cycle of clock operation, i.e. $k=1,2,\ldots$. After each cycle, the measurement outcome $m_k$ providing the phase estimate $\phi_{\rm est}(m_k)$ is used to infer an estimated frequency deviation $\phi_{\rm est}(m_k)/T$. In combination with previous measurement results this is used to correct the laser frequency fluctuations via a feedback loop yielding the corrected frequency of the clock $\omega(t)$. For further details on the actual clock operation we refer to App.~\ref{app:Clock_simulation}, where we also describe our numerical simulations of optical atomic clocks. We emphasize the importance of finite dynamic range in phase estimation in identifying the optimal clock operation, as provided by the Bayesian approach of Sec.~\ref{sec:interferometer}.

The relevant quantity characterizing the long-term clock instability is the Allan deviation $\sigma_y(\tau)$ for fluctuations of fractional frequency deviations $y\equiv[\omega(t)-\omega_A ]/\omega_A $, averaged over time $\tau\gg  T$~\cite{Ludlow2015}. 
To connect the Bayesian posterior phase variance of the optimized interferometer~\eqref{eq:BMSQE} of Sec.~\ref{sec:interferometer}, we follow the approach of~\cite{Leroux2017} to obtain predictions for the clock instability in the limit of large averaging time $\tau$. Our predictions are supported by numerical simulations of the closed servo-loop of the optical atomic clocks. 

In the following we assume that interrogation cycles can be performed without dead times (Dick effect). This can be achieved using interleaved interrogation of two ensembles~\cite{Schioppo2017}. For interrogation of a single ensemble, Dick noise may pose limitations for interaction-enhanced protocols especially for larger ensembles, as was analyzed for squeezed states in Ref.~\cite{Schulte2020b}.  In App.~\ref{app:dead_time} and App.~\ref{app:cum_int} we characterize in more detail the requirements regarding dead time for the class of variational protocols developed here.

\subsection{Prediction of clock instability in the Bayesian framework}\label{subsec:PredictionClock}

\begin{figure}[t]
    \centering
    \includegraphics[width=\columnwidth]{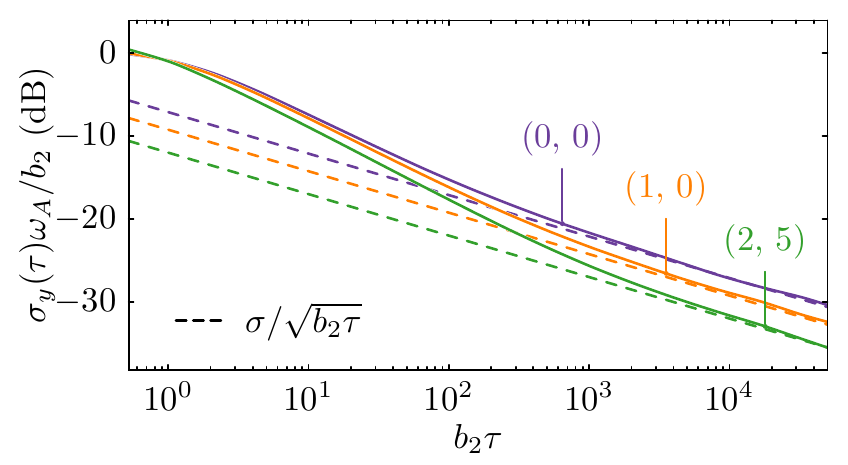}
    \caption{Allan deviation representing a single run of a flicker frequency noise limited clock servo loop based on variationally optimized $N = 64$ particle interferometers at different circuit complexities $(n_{\rm En}, n_{\rm De})$ and a Ramsey interrogation time  $b_2 T \approx 0.5$ (solid lines). For comparison the long time scaling predicted by Eq.~\eqref{eq:Allan_dimless} is shown by the dashed lines. }
    \label{fig:AllanDeviation_Run}
\end{figure}

As shown in~\cite{Leroux2017}, the Allan deviation can be well approximated by means of the effective measurement uncertainty $\Delta\phi_M$ which isolates the measurement contribution from the prior knowledge, as in Eq.~\eqref{eq:effective_measurement_variance}. Assuming no dark times between interrogation cycles, the Allan deviation reads
\begin{equation}
\sigma_{y}(\tau) = \frac{1}{\omega_A } \frac{\Delta\phi_M( T)}{T } \sqrt{\frac{T }{\tau}}.
\label{eq:Allan}
\end{equation}
Here $\tau/ T$ is the number of cycles of clock operation and $\Delta\phi_M(T) \equiv [(\Delta\phi_{T})^{-2}-(\delta\phi_{T})^{-2}]^{-1/2}$ is the effective measurement uncertainty of one cycle. The posterior width $\Delta\phi_T$ is found according to~\eqref{eq:BMSQE} assuming a prior width $\delta\phi_{T}=(b_{\alpha} T)^{\alpha/2}$ corresponding to laser noise dominated spreading of the phase distribution within one interrogation cycle. The labels $\alpha = 1, 2, 3$ specify temporal correlations in the phase noise of the laser and correspond to atomic clocks with a white-, \mbox{flicker-,} or random-walk-frequency-noise-limited laser, respectively. The laser noise bandwidth $b_{\alpha}$ and the exponent $\alpha$ are related to the power spectral density $S_{L}(f) \propto f ^{1-\alpha}$ of the free running laser (see App.~\ref{sec:PSD}). Representative examples for $\sigma_y (\tau)$ when using variationally optimized protocols are shown in Fig.~\ref{fig:AllanDeviation_Run}. The solid lines result from numerical simulations of the full feedback loop of an atomic clock in which an integrating servo corrects out frequency fluctuations over the course of multiple cycles, see App.~\ref{app:Clock_simulation} for details. For the simulations we assume the atoms as ideal frequency references without any systematic shift of $\omega_A$. 

\begin{figure*}[t]
    \centering
    \includegraphics[width=\columnwidth]{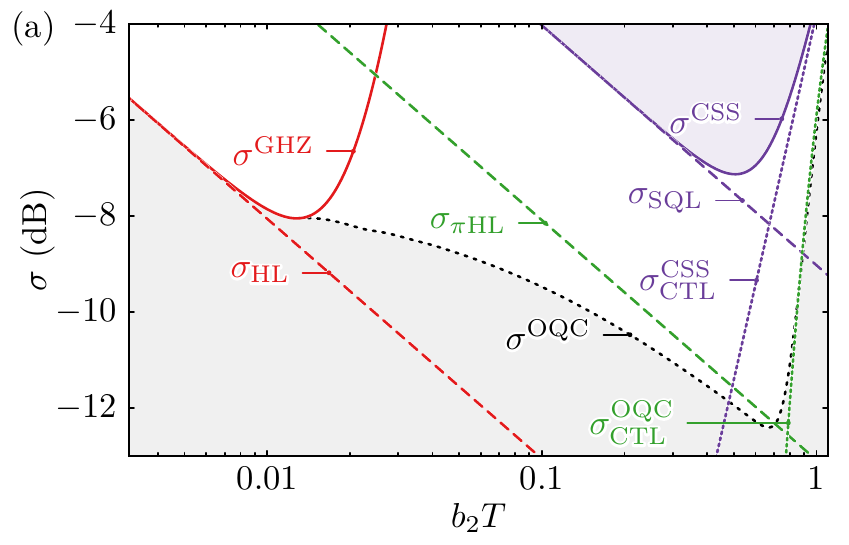}
    \includegraphics[width=\columnwidth]{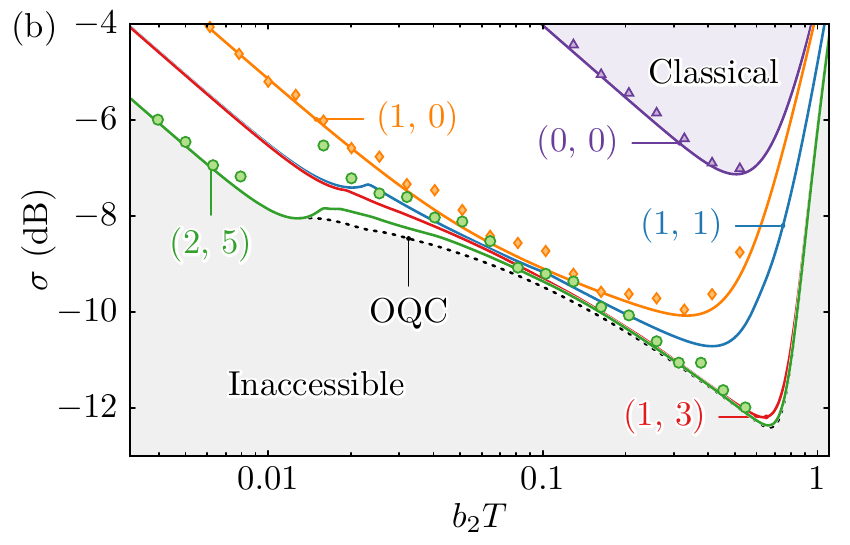}
    \caption{Dimensionless Allan deviation $\sigma$ [see Eq.~(\ref{eq:Allan_dimless})] of a $N=64$ flicker-frequency-noise limited clock at constant averaging time $\tau$ as a function of the Ramsey interrogation time $T$ rescaled to the bandwidth of the laser noise. The dotted black line indicates the instability of the optimal quantum clock (OQC). (a) Analytic expressions for the instability of coherent spin state (CSS) and GHZ-state clock (solid lines) in comparison to quantum projection noise limits, namely the standard quantum noise limit (SQL), $\pi-$corrected Heisenberg limit ($\pi$HL),  HL (dashed) and the coherence time limits (CTL) of a CSS clock and the OQC (densely dotted). (b) Variational approximation of the OQC at increasing circuit complexities $(n_{\rm En}, n_{\rm De})$ with $3(n_{\rm En}+ n_{\rm De})$ variational parameters. Markers show the numerically determined instability extracted from simulations of the full feedback loop of an atomic clock, described in detail in App.~\ref{app:Clock_simulation}, for a selection of optimized protocols. The numerical data is displayed only up to values of $b_{2} T$ where no fringe-hop occurred within the $2\times 10^6$ simulated clock cycles. Beyond this point an abrupt loss of stability was observed.}
    \label{fig:AllanDeviation_hT}
\end{figure*}

In atomic clocks the simulated Allan deviations presented in Fig.~\ref{fig:AllanDeviation_Run} are larger at small averaging times $\tau/T \sim 1$, due to the delayed feedback, before reducing as $\sigma_y(\tau)\propto\tau^{-1/2}$ at long averaging times $ \tau/T \gg 1$ when all correlated laser noise is corrected out.
To determine long-term stability the Allan deviation is measured experimentally for a time $\tau$ long enough that clock instability has reached this asymptotic scaling. Therefore, we introduce and consider below a dimensionless prefactor for the asymptotic scaling
\begin{align}
    \sigma = \frac{\Delta\phi_M( T)}{\sqrt{b_{\alpha}  T}},
    \label{eq:Allan_dimless}
\end{align}
which gives the Allan deviation in units of $\smash{\omega_A ^{-1}(b_{\alpha}/\tau)^{1/2}}$, as shown by the dashed lines in Fig.~\ref{fig:AllanDeviation_Run}. In the following, we use Eq.~\eqref{eq:Allan_dimless} to re-evaluate the performance of the optimized interferometers presented in Fig.~\ref{fig:results} as the achievable long-term clock instability $\sigma$ at an averaging time $\tau$. In comparison to the framework of Sec.~\ref{sec:interferometer} the BMSE is replaced by the Allan deviation and the prior width by the interrogation time $T$. We note that the scaling of the Allan deviation with respect to $T$ is more intricate than the one of the BMSE with the prior width: On the one hand, a large interrogation time means good accuracy in frequency estimation, but on the other hand, it also broadens the prior distribution and therefore degrades the phase estimation.

\subsection{Results of the clock optimization}

Figure~\ref{fig:AllanDeviation_hT}(a,b) shows the achievable long-term clock instability $\sigma$ as a function of the interrogation time $ T$ for clocks made of $N=64$ atoms and the flicker-noise-limited laser.
The purple line (in both panels) represents performance of the conventional clock exploiting Ramsey interferometer with CSS as input, collective spin projection measurement, and a linear estimator, given by the circuit $(0,0)$. Thus, the shaded area above the purple line roughly defines the performance achievable by classical clocks. In the case of CSS based classical clocks the cost function~\eqref{eq:BMSQE} can be analytically minimized~\cite{Leroux2017} yielding the dimensionless Allan deviation
\begin{equation}
\sigma^{\rm CSS} = \sqrt{\frac{1}{b_{\alpha} T}\left[\frac{e^{\nu}}{N}+\Big(1-\frac1N\Big)\mathrm{sinh}\,\nu-\nu\right]},
\label{eq:Allan_CSS}
\end{equation}
where $\nu \equiv (\delta\phi_{T})^2$. The expression~\eqref{eq:Allan_CSS} has two important limits. For small interrogation times and, consequently, small prior widths the performance of the clock is limited by the quantum projection noise of the uncorrelated atoms as $\sigma_{\rm SQL} = (N\, b_{\alpha} T)^{-1/2}$. The SQL limited clock instability $\sigma_{\rm SQL}$ (dashed purple line) decreases as the interrogation time grows. For large interrogation times, $b_{\alpha}  T\sim1$, however, the laser noise becomes dominant and generates accumulated phase values exceeding the dynamic range of the atomic interferometer, thus, leading to the laser coherence time limit (CTL)~\cite{Schulte2020b} of the clock $\sigma^{\rm CSS}_{\rm CTL} = \{{[\mathrm{sinh}(\delta\phi_T^2)-\delta\phi_{T}^2]}/(b_{\alpha} T)\}^{1/2}$ (dotted purple line). Between these two limits there exists an optimal interrogation time delivering the minimum Allan deviation $\sigma_{\rm opt}\equiv\min_{ T}\sigma$ which defines the optimal clock performance.

The black dotted line in Fig.~\ref{fig:AllanDeviation_hT}(a,b) shows the instability of the optimal quantum clock (OQC), $\sigma^{\rm OQC}$, exploiting single-shot protocols with the optimal interferometer. The gray shaded region below the black dotted curve is inaccessible to any $N$-particle clock not using entanglement between different clock cycles for initial state preparations and/or measurements. The laser CTL for the optimal clock in the asymptotic limit of large $N$ can be estimated from Eq.~\eqref{eq:BMSQE_POVM} by assuming zero phase estimation error within the $[-\pi,\pi]$ interval and $\epsilon(\phi)=4\pi^2$ outside of the interval due to the phase slip
\begin{equation}
\sigma_{\rm CTL}^{\rm OQC}= \sqrt{\frac{4\pi^2}{b_{\alpha} T}\Big(1-\mathrm{erf}\frac{\pi}{\sqrt{2}\,\delta\phi_{ T}}\Big)}.
\label{eq:CTL_OQC}
\end{equation}
The green dotted line in panel (a) shows the laser CTL for the optimal clock, $\sigma_{\rm CTL}^{\rm OQC}$. The optimal clock instability at shorter interrogation times demonstrates two distinct scalings corresponding to the two Heisenberg limits discussed in Sec.~\ref{subsec:HL}. At very short times, $(b_{\alpha}  T)^{\alpha/2}\lesssim N^{-1}$, the GHZ state based clock (red line) becomes optimal approaching the instability limit given by the conventional HL, $\smash{\sigma_{\rm HL} = N^{-1}({b_{\alpha} T})^{-1/2}}$ (red dashed line). Larger interrogation times correspond to wider prior phase distributions hence the $\pi$-corrected HL becomes the limiting factor, $\smash{\sigma_{\pi{\rm HL} } = \pi N^{-1}({b_{\alpha} T})^{-1/2}}$~(green dashed line). The optimal quantum clock instability in the limit of large number of atoms, $N\to\infty$, is fundamentally restricted by the interplay between the $\sigma_{\pi{\rm HL} }$ and $\sigma^{\rm OQC}_{\rm CTL}$ as we will discuss below.

The instabilities of clocks based on variationally optimized interferometers employing quantum circuits of various complexities are shown in Fig.~\ref{fig:AllanDeviation_hT}(b) with solid color lines. In particular, the orange line corresponds to the SSS based clock, given by the circuit $(1,0)$. As the circuits depth grows, the enhanced dynamic range of the variational interferometer shifts the laser CTL towards larger interrogation times which in combination with suppressed shot noise reduces the clock instability. The figure shows that variational clocks of growing complexity quickly outperform the SSS clock and approach the optimal quantum clock instability. 
Beyond the model predictions this improvement is also observed in simulations of a full clock operation using variationally optimized protocols, as shown by the markers in Fig.~\ref{fig:AllanDeviation_hT}(b). Deviations between theory and numerical results can arise due to a number of different effects. For one, the onset of fringe-hops for $b_2 T \sim 1$ is not included explicitly in the models. Especially for small $N$ a sudden loss of stability, resulting from fringe-hops, can occur before reaching the CTL due to stronger, non-Gaussian measurement noise~\cite{Leroux2017, Schulte2020b}. In contrast, for clocks with larger $N$ and increasing complexity it is expected that the onset of fringe-hops and the minimum of CTL coincide.
Another source of discrepancy is the assumption of a laser noise dominated prior width $\delta\phi_{T}=(b_{\alpha} T)^{\alpha/2}$. Propagation of the measurement uncertainty and delay within the feedback control can lead to a broadening of the true phase distribution.
Especially protocols which are highly optimized to a particular prior width may thus not achieve their predicted stability in the simulations, e.g. around $b_2 T \approx 0.02$ in Fig.~\ref{fig:AllanDeviation_hT}(b).

Nevertheless, good agreement between the numerically determined instability and the theory prediction is found around the overall optimal protocols.

\begin{figure}[t]
\centering
\includegraphics[]{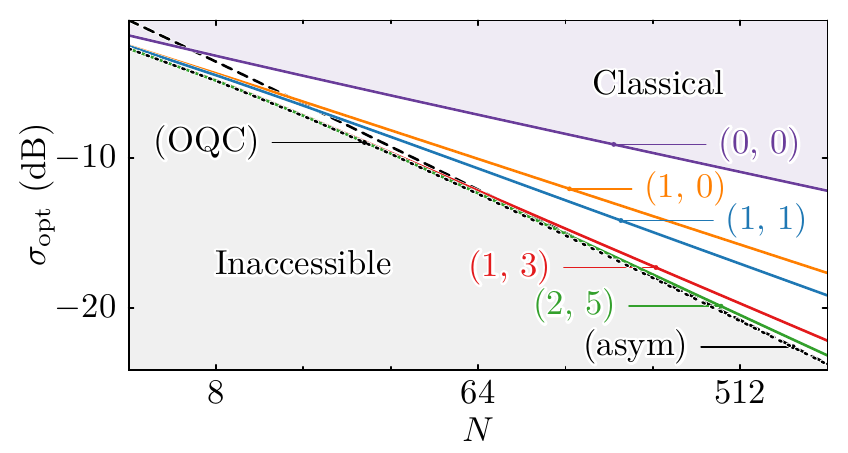}
\caption{Particle number $N$ dependence of the dimensionless Allan deviation of a flicker-frequency-noise limited clock at a constant averaging time $\tau$ and the optimal Ramsey interrogation time for different circuit complexities $(n_{\rm En}, n_{\rm De})$ with $3(n_{\rm En}+ n_{\rm De})$ variational parameters. For comparison we also show the performance of the optimal quantum clock (dotted line) and the asymptotic [see Eq.~\eqref{eq:ScalingOQC}] obtained from the $\pi$-corrected HL  (dashed line). }
\label{fig:AllanDeviation_N}
\end{figure}

In Fig.~\ref{fig:AllanDeviation_N} we study optimal instability of the variational clocks $\sigma_{\rm opt}$ (corresponds to minima in Fig.~\ref{fig:AllanDeviation_hT})  as a function of the atomic ensemble size $N$. The CSS clock is represented by the purple line which scales asymptotically as $\sigma_{\rm opt}^{\rm CSS}\propto N^{-(3\alpha-1)/(6\alpha)}$. The scaling is a bit slower than the conventional SQL limit $\smash{\propto N^{-1/2}}$ due to the laser CTL which reduces the optimal interrogation time as $N$ grows. Any classical clock using one-shot protocols with collective spin measurements belongs to the shaded purple region above the CSS clock line. 

The $N$-scaling of the optimal quantum clock is shown with the black dotted line for system sizes up to $N=64$. For larger system sizes we show the asymptotic behavior (black dashed line) obtained by combining the noise contributions of the $\pi$-corrected HL and the laser CTL, $\smash{\sigma_{\rm asym}\equiv\min_{ T}[{\sigma_{\pi\rm HL}^2+(\sigma_{\rm CTL}^{\rm OQC })^2}]^{1/2}}$. Similar to the classical clock scaling, the laser CTL prevents the optimal quantum clock (OQC) from achieving the Heisenberg scaling $\smash{\propto N^{-1}}$, instead, leading to a logarithmic correction in the large $N$ limit as found in~\cite{Borregaard2013a,Kessler2014}. The present approach allows obtaining tighter bounds on the asymptotic scaling for general $\alpha$ (see App.~\ref{app:N_scaling}). In particular, for the flicker-noise-limited laser, $\alpha=2$, the OQC instability scales as
\begin{equation}
\sigma_{\rm opt}^{\rm OQC} \propto \frac{\sqrt{\pi}}{N}\left[\ln (z \ln z)^{-1/2}+\ln (z \ln z)^{1/2}\right]^{1/2},
\label{eq:ScalingOQC}
\end{equation}
with $z\equiv {32N^4}/{\pi}$ and the corresponding optimal interrogation time scaling as $T_{\rm opt}^{\rm OQC}\simeq \pi b^{-1}_{2}\ln(z \ln z)^{-1/2}$.
The gray shaded area below the dashed and dotted black lines is inaccessible to quantum clocks without entangled clock cycles.
Finally, the variationally optimized clocks of various circuit complexities are shown with solid color lines and demonstrate scalings approaching the optimal quantum clock as the circuit depth increases.

We have also studied performance of the variationally optimized clocks experiencing individual atomic dephasing during the interrogation period $T$. Similar to the results of Sec.~\ref{subsec:dephasing}, the optimized clocks perform well for decoherence rates small compared to the laser noise bandwidth, $\gamma/b_{\alpha}\ll1$. For stronger noise, $\gamma/b_{\alpha}\gtrsim1$, the optimized clock instability approaches the one of the classical clock, as expected. We also checked the performance of optimized clocks for other types of laser noise $\alpha = 1,3$, and found no significant changes to the results presented above.

In summary, atomic clocks based on variational quantum interferometers with low-depth circuits can approach the performance of the optimal quantum clock in single-shot protocols. The variationally optimized clocks can be readily complemented with more sophisticated interrogation schemes~\cite{Borregaard2013,Rosenband2013}, eventually also approaching the ultimate quantum bound on the Allan deviation~\cite{Chabuda2016,Chabuda2020}.

\section{Outlook and Conclusions}

In this work we have studied optimal Ramsey interferometry for phase estimation with entangled $N$-atom ensembles, and application of these optimal protocols to atomic clocks. We have considered a Bayesian approach to quantum interferometry, and have defined optimality via a cost function, which in the present study is the BMSE for a given prior distribution or, in the context of atomic clocks, the Allan deviation for a given Ramsey time. The key feature of the present work is that optimization is performed within the family of operational quantum resources provided by a particular programmable quantum sensor platform. Thus identifying the optimal quantum sensor is recast as a variational quantum optimization where the entangling circuits generating the optimal input state, and the decoding circuits implementing the optimal generalized measurement are variationally approximated with the given resource up to a certain circuit depth. We have presented two model studies: in our  first model, we considered one-axis twisting as quantum resource; our second model uses finite range interactions as entangling operations. Our examples demonstrate that already low-depth circuits provide excellent approximations for optimal quantum interferometry. We emphasize that the familiar discussions of interferometry with spin-squeezing and GHZ states are included as special cases. Furthermore, advanced measurement strategies including adaptive measurement and quantum phase estimation are not advantageous for the present problem, as a von Neumann measurement has been proven optimal.

Given advances in building small atomic scale quantum computers, or programmable quantum simulators which can act also as quantum sensors, the variational approach to optimal quantum sensing provides a viable route to entanglement enhanced quantum measurements with existing experimental entangling, possibly non-universal resources, and optimizing in presence of noise. Indeed trapped ions with M{\o}lmer-S{\o}rensen entangling gates, and optical arrays interacting via Rydberg finite range interactions or cavity setups provide the necessary ingredients for implementing such variational protocols, and quantum sensors. While first generation experiments might demonstrate optimal Ramsey interferometry for a specified dynamic range of the phase, and optimization of quantum circuits `on the quantum sensor' for various circuit depths (Sec.~\ref{sec:interferometer}), the present work also promises application of variational quantum sensing on existing quantum sensors, in particular atomic clocks (Sec.~\ref{sec:clock}). The guiding principle behind the present work of identifying for a sensing task the optimal sensing protocol given the quantum resources provided by a particular sensor and sensor platform, is, of course, general and generic, and applies beyond Ramsey interferometry, and beyond the BMSE as cost function.

As an outlook, we emphasize that the search for optimal sensing can also be run directly as a  quantum--classical feedback loop on the physical quantum sensor. This offers the intriguing possibility of optimizing with given quantum resources and in presence of imperfections of the actual device, which might include control errors and noise. Further studies are needed to explore best optimization strategies of the cost function on the classical side of the optimization loop given the limited measurement budget on the programmable quantum sensor. This applies to both the initial global parameter search, supported by theoretical modeling, and small iterative readjustments of optimal operation points due to slow drifts of the quantum sensor. 

Optimization \textit{on} the (physical) quantum sensor can also be performed in the regime of large particle numbers $N$, which might be inaccessible to classical computations, i.e.~in the regime of quantum advantage. Hybrid classical-quantum algorithms have been discussed previously as variational quantum eigensolvers for quantum chemistry and quantum simulation, where  `lowest energy' plays the role of the cost function which is evaluated on the quantum device. In contrast, in variational quantum sensing we optimize quantum circuits in view of an `optimal measurement' cost function, and it is the (potentially large scale) entanglement represented by the variational many-particle wavefunction in $N$-atom quantum memory, which provides the quantum resource and gain for the quantum measurement.

\emph{Note added.} After submission of the present manuscript, Ref.~\cite{Marciniak2021} reported an experimental implementation of  variationally optimized Ramsey interferometry in a systems of up to $N=26$ trapped ions,  in one-to-one correspondence to the present theoretical work. This includes demonstration of quantum enhancement in metrology beyond squeezing through low-depth, variational quantum circuits, and on-device quantum-classical feedback optimization to `self-calibrate' the variational parameters. In both cases it is found that variational circuits outperform classical and direct spin squeezing strategies under realistic noise and imperfections.

\begin{acknowledgments} We thank R.~Blatt,  T.~Feldker, A.~Kaufman, D.~Leibrandt, K.~Macieszczak,  C.~Marciniak, T.~Monz,  I.~Pogorelov, P.~Schmidt, P.~Silvi and Jun Ye for discussions and valuable comments on the manuscript. Computational results were based on the LEO HPC infrastructure of the University of Innsbruck.  Research  in  Innsbruck  is supported by the US Air Force Office of Scientific Research (AFOSR) via IOE Grant No.~FA9550-19-1-7044 LASCEM, the European Union’s Horizon 2020 research and innovation program under Grant Agreement No. 817482 (PASQuanS) and No. 731473 (QuantERA via QTFLAG), and by the Simons Collaboration on Ultra-Quantum Matter, which is a grant from the Simons Foundation (651440, P.Z.), and by the Institut f\"ur Quanteninformation. Innsbruck theory is a member of the NSF Quantum Leap Challenge Institute Q-Sense. M.S. and K.H. acknowledge funding by the Deutsche Forschungsgemeinschaft (DFG, German Research Foundation) under Germany's Excellence Strategy – EXC-2123 QuantumFrontiers – 390837967 and CRC 1227 `DQ-mat' project A06.
\end{acknowledgments}

\appendix 

\section{Laser noise and prior distribution width}\label{sec:PSD}
To present results in Sec.~\ref{sec:clock} in dimensionless units, we follow~\cite{Leroux2017} and define an effective bandwidth $\tilde{b}$ via 
\begin{align}
    \sigma_{L}(1/\tilde{b}) \ \omega_{A} /\tilde{b} = 1,
    \label{eq:effective_bandwidth}
\end{align}
where $\sigma_{L}$ is the Allan deviation of the uncorrected reference laser.  
For a laser that is mainly limited by a single power spectral density component, i.e $S_{L}(f) = h_{1-\alpha} f^{1-\alpha}$ one can unambiguously express the bandwidth in terms of the prefactors $h_{1-\alpha}$ in the power spectral density and the respective Allan deviation~\cite{Barnes1971}, so that
\begin{align*}
    \tilde{b}_{1} &= \frac{h_0}{2}\omega_{A}^{2}, \notag \\
    \tilde{b}_{2} &= \sqrt{h_{-1}2\ln 2}~ \omega_{A}, \notag \\
    \tilde{b}_{3} &= (h_{-2}/6)^{-1/3} (2\pi)^{-2/3}\omega_{A}^{-2/3}.
\end{align*}
Numerical simulation of the clock feedback loop~\cite{Leroux2017} reveal that the dimensionless time $b_{\alpha}T$ is related to the prior distribution width of a stabilized clock by the relation 
$
    (\delta \phi) ^2 = (b_{\alpha} T)^{\alpha},
$
where $b_{\alpha} = \chi(\alpha)^{1/\alpha} \tilde{b}_{\alpha} $ is a rescaled bandwidth, differing from $\tilde{b}_{\alpha}$ only by an empirically determined prefactor $\chi \approx 1,1.8,2$ for $\alpha=1,2,3$.
For a laser spectrum containing all three contributions Eq.~\eqref{eq:effective_bandwidth} can still be used to determine an effective bandwidth, and servo loop simulations of the clock can reveal the modified time dependence of the prior distribution width enabling one to extend the clock model to realistic laser noise parameters.

\section{Spin $x$-parity in entangling and decoding circuits}\label{sec:spin_x_parity}
We consider global rotations $\mathcal{R}_{\mu}$, OAT interactions $\mathcal{T}_{\mu}$ (see Sec.~\ref{subsec:VariationalInterferometer}) and finite range dressing interactions $\mathcal{D}_{\mu}$ (see Sec.~\ref{sec:FiniteRange}) with $\mu=x,y,x$ as resources for the variational optimization. Within is this set of resources we are able to ensure an anti-symmetric estimator by imposing invariance under the spin $x$-parity $P_x$ on the Entangler and Decoder, i.e. $P_x \mathcal{U}_{\rm En}\mathcal{R}_y(-\pi/2)P_x= \mathcal{U}_{\rm En}\mathcal{R}_y(-\pi/2)$ and $P_x \mathcal{U}_{\rm De}P_x=\mathcal{U}_{\rm De}$ under the  spin $x$-parity $P_x = \mathcal{R}_x(\pi/ 2)$, since this implies 
\begin{align}
    \bar \phi_{\rm est}(\phi)&=\bra{\psi_0}  \mathcal{U}_{\rm En}^{\dag} e^{i \phi J_z} \mathcal{U}_{\rm De}^{\dag} J_y \mathcal{U}_{\rm De} e^{- i \phi J_z} \mathcal{U}_{\rm En}\ket{\psi_0} \notag \\ &= -\bra{\psi_0} \mathcal{U}_{\rm En}^{\dag}  e^{- i \phi J_z}  \mathcal{U}_{\rm De}^{\dag} J_y \mathcal{U}_{\rm De} e^{i \phi J_z} \mathcal{U}_{\rm En}\ket{\psi_0}\notag \\
    & = - \bar \phi_{\rm est}(-\phi)
\end{align}
where we use that $P_x J_x P_x = J_x$, $P_x J_{y,z}P_x = -J_{y,z}$, $P_x^{\dagger} = P_x$  and  $ P_x \mathcal{R}_y(\pi/2) \ket{\psi_0} = \mathcal{R}_y(\pi/2) \ket{\psi_0}$. The most general entangling and decoding sequences satisfying these constraints are used in Eq.~\eqref{eq:Entangler},\eqref{eq:Decoder} and displayed in Fig.~\ref{fig:Fig1}. 

\section{Wigner distribution}\label{sec:WignerDistribution}
In Secs.~\ref{sec:ResultsOptimization}, \ref{subsec:dephasing} we visualize collective spin operators $\mathcal O$ like the density matrix $\rho=\ket{\psi_{\rm in}}\bra{\psi_{\rm in}}$ of the initial state of the interferometer [Eq.~\eqref{eq:Psi_in}], or the variationally decoded measurement operator $M=\sum_{m}\phi_{\rm est}(m)  \mathcal{U}_{\rm De}^{\dagger}\ket{m}\bra{m}\mathcal{U}_{\rm De}$ [decomposed in terms of the projection in Eq.~(\ref{eq:pi})] by means of the Wigner distribution~\cite{Dowling1994}. 

To obtain the Wigner distribution, the operator is expanded in terms of spherical tensors 
\begin{align}
T_{k,q}=\sum_{m,m'=-N/2}^{N/2} (-1)^{j-m} \sqrt{2k+1} \quad \notag \\
\times\ \bigl(\begin{smallmatrix}j & k & j\\ -m & q & m' \end{smallmatrix}\bigr) \ket{m}\bra{m'},
\end{align}
where $\bigl(\begin{smallmatrix}j & k & j\\ -m & q & m' \end{smallmatrix}\bigr)$ denotes the Wigner $3j$ symbol. $\mathcal{O}$ can be represented in the spherical tensor basis 
\begin{align}
    \mathcal O =\sum_{k=0}^{N}\sum_{q=-k}^{k}c_{k, q} T_{k,q}
\end{align} where $c_{k,q}=\tr\big(\mathcal OT_{k,q}\big)$.
Replacing $T_{k,q}$ in this representation by spherical harmonics $Y_{k,q}(\theta, \phi)$, one arrives at the Wigner distribution,
\begin{align}
W_{\mathcal O}(\theta, \phi) = \sum_{k=0}^{N}\sum_{q=-k}^{k}c_{k,q}Y_{k,q}(\theta, \phi)
\end{align}
as a quasi-probability distribution on a generalized Bloch sphere. 

The Wigner function can be used to calculate the expectation value 
\begin{align}
    \tr(\rho M)=\int_{0}^{\pi}d \theta \int_{0}^{2 \pi}d \varphi W_{M}(\theta, \varphi)W_{\rho}(\theta, \varphi)
\end{align}
by integrating the overlap of the respective Wigner functions over the generalized Bloch sphere. This implies that we can interpret contours of the measurement distribution with the different eigenvalues of the measurement operator while the amplitude of the state distribution indicates how much the state overlaps with the respective projection of the measurement projection.

\section{Numerical optimization of the phase-operator based interferometer}
\label{app:covarinat_measurement}

Here we define the phase operator and describe an iterative optimization procedure allowing us to minimize the cost function~\eqref{eq:BMSQE_POVM} for a given observable using the Minimal Mean Squared Error (MMSE) estimator~\cite{Demkowicz2015}. The phase operator $\hat\Phi$ reads~\cite{Derka1998,Pegg_1988}:
\begin{align}
\label{eq:phase_operator}
\hat{\Phi}&=\sum_{s=-J}^{J}\phi_{s}\ket s\bra s,\\
\phi_{s}&=\frac{2\pi s}{2J+1},\\
\label{eq:phase_eigenstate}
\ket s&=\frac{1}{\sqrt{2J+1}}\sum_{m=-J}^{J}e^{-i\phi_{s}m}\ket m,
\end{align}
where $J_{z}\ket m=m\ket m$.

Our goal is to minimize the cost function Eq.~\eqref{eq:BMSQE_POVM} for the observable $\hat\Phi$ and the MMSE estimator by finding the optimal initial state $\ket{\psi_{\hat\Phi}}$. The MMSE estimator reads~\cite{Demkowicz2015}:
\begin{equation}
\label{eq:MMSE_estimator}
\phi_{\rm est}^{\rm MMSE}(s)=\int\phi\,p(\phi|s)d\phi,
\end{equation}
where the conditional probability is $p(\phi|s)\propto p(s|\phi)\mathcal{P}(\phi)$ with $p(s|\phi)=|\braket{s|e^{-i\phi J_z}|\psi_{\rm in}}|^2$ and the observable eigenstate $\ket{s}$ defined in Eq.~\eqref{eq:phase_eigenstate}. 

The optimization is performed iteratively. Initially we start with $s=0$ eigenstate of $\hat\Phi$ as the input state $\ket{\psi_{\rm in}^{(0)}}=\ket{s=0}$, which is a good approximation for a state highly sensitive to phases around $\phi=0$. The state defines the corresponding MMSE estimator $\phi_{\rm est (0)}^{\rm MMSE}(s)$ as given by Eq.~\eqref{eq:MMSE_estimator}. In the next iteration we find the state $\ket{\psi_{\rm in}^{(1)}}$ minimizing the cost function~\eqref{eq:BMSQE_POVM} for the given $\phi_{\rm est (0)}^{\rm MMSE}(s)$ estimator by solving a corresponding eigenproblem, as described in~\cite{Macieszczak2014}. The iterative procedure converges quickly yielding the optimal initial state for the POI $\ket{\psi_{\rm in}^{(k)}}\to_{k\to\infty}\ket{\psi_{\hat\Phi}}$ which, in turn, defines the optimal estimator via Eq.~\eqref{eq:MMSE_estimator} and the corresponding posterior width $\Delta\phi_{\rm POI}$. This result is used in Sec.~\ref{subsec:covariant_measurement_comparison}.

\section{$N$-scaling of the optimal quantum clock instability}\label{app:N_scaling}
Here we derive asymptotic scaling of the optimal interrogation time and the corresponding minimal instability of the optimal quantum clock. As discussed in Sec.~\ref{sec:clock}, the instability of clocks exploiting single-shot protocols is fundamentally limited by the measurement shot noise given by the $\pi$-corrected HL for short interrogation times $T$, and the laser CTL for large $T$. For the dimensionless Allan variance we write 
\begin{align}
(\sigma^{\rm OQC})^2 &=\sigma_{\pi\rm HL}^2 + (\sigma_{\rm CTL}^{\rm OQC })^2,\notag\\
&= \frac1{s\pi^{\frac2\alpha}}\left\{ \frac{\pi^2}{N^2} +
4\pi^2\left[1-\mathrm{erf}\big(\frac1{\sqrt{2s^{\alpha}}}\big)\right]\right\}.
\label{eq:app:instability}
\end{align}
where $s\equiv\pi^{-\frac2\alpha}b_{\alpha} T$ is the dimensionless Ramsey time. The goal is to minimize Eq.~\eqref{eq:app:instability} with respect to $s$ in the limit of large number of atoms, $N\to\infty$. The derivative with respect to $s$ reads
\begin{align*}
\frac{d}{ds}(\sigma^{\rm OQC})^2 = -\frac1{s^2\pi^{\frac4\alpha}}\Big\{ &\frac{\pi^2}{N^2} +
4\pi^2\left[1-\mathrm{erf}\big(\frac1{\sqrt{2s^{\alpha}}}\big)\right]\\
& -
2\sqrt{\frac{2\pi^3\alpha^2}{s^\alpha}} e^{-\frac1{2s^{\alpha}}}
\Big\},
\end{align*}
and, using a self-consistent assumption for optimal time $s_{*}\ll1$, results in the following equation for $s_{*}$,
\begin{equation}
e^{s_*^{-\alpha}}s_*^{\alpha} = \frac{8\alpha^2 N^4}{\pi}.
\label{eq:app:s_opt}
\end{equation}
Here we used the error function asymptotic $1-\mathrm{erf}(x)\to e^{-x^2}/(\sqrt\pi x)$ for $x\to\infty$.
Taking the logarithm of the expression~\eqref{eq:app:s_opt} ($s_*$, $\alpha$, and $N$ are positive) we obtain an equation for $w\equiv s_*^{-\alpha}$,
\begin{equation*}
w - \ln w = \ln z,
\end{equation*}
with $z\equiv 8\alpha^2 N^4/\pi$. For $z>e$, the solution can be written as the infinitely nested logarithm, $w(z) = \ln(z\ln(z\ln(z(\ln\ldots)\ldots)))$, and can be checked by direct substitution. Using the $w(z)$ function we can express the optimal Ramsey time for $N\gg1$ as follows
\begin{align} 
b_{\alpha} T_{\rm opt} = \pi^{\frac2\alpha}s_*
&= \left[\frac1{\pi^2} w(z)\right]^{-\frac1\alpha}
\label{eq:app:T_opt}
\end{align}
Finally, we substitute the optimal Ramsey time into Eq.~\eqref{eq:app:instability}
\begin{equation}
(\sigma^{\rm OQC}_{\rm opt})^2 \simeq \frac{\pi^2}{N^2}\left[\frac{w(z)}{\pi^2}\right]^{\frac1\alpha}\left[ 1 + \frac{2}{\alpha w(z)}\right].
\label{eq:app:sigma_opt}
\end{equation}
We use Eqs.~\eqref{eq:app:T_opt} and \eqref{eq:app:sigma_opt} and keep only the first two logarithms in the definition of $w(z)$ to obtain expressions for the optimal interrogation time and minimal instability of the optimal quantum clocks in Sec.~\ref{sec:clock} for $\alpha=2$.

\section{Finite dead time in the atomic clock protocol}\label{app:dead_time}
Here we discuss upper limits to the dead times of atomic clocks, which are required to reach the variationally optimized stability presented in Sec.~\ref{sec:clock}.
When each interrogation cycle of duration $T_{C} = T_{D} + T$ is composed of a dead time $T_{D}>0$, and  Ramsey free evolution time $T$, the stability is reduced compared to the ideal case at $T_{D}=0$ discussed in the main text. 

Let us consider $S_L (f) = h_{-1} f^{-1}$ as the power spectral density of the free running laser. In addition, we assume that the protocols are sensitive to phase shifts during $T$ only and that all entangling and decoding operations are included in the dead time where we assume no sensitivity.
Given these assumptions, the instability contribution of the Dick effect is~\cite{Dick_local_1987}
\begin{equation}
    \sigma_{\rm Dick}^2 (\tau) = \frac{b_2}{\omega_A^2 \tau} \frac{b_2 T}{\chi(2) 2 \ln 2} \frac{1}{d^3} \sum_{n=1}^{\infty} \frac{\sin^2 (\pi n d)}{\pi^2 n^3}
\end{equation}
with $\chi$ given in App.~\ref{sec:PSD} and the duty cycle $d=T/T_{C}$. In addition, the instability predicted in the Bayesian framework, Eq.~\eqref{eq:Allan}, becomes
\begin{equation}
    \sigma_{\rm Bay}^2(\tau) = \frac{b_2}{\omega_A^2 \tau} \frac{\sigma^2}{d}
\end{equation}
with $\sigma$ as defined in Eq.~\eqref{eq:Allan_dimless}.

In the following we want to estimate below which level of dead time the combined instability $\sigma_y (\tau) = \sqrt{\sigma_{\rm Bay}^2(\tau) + \sigma_{\rm Dick}^2 (\tau)}$ is no longer dominated by the contribution of the Dick effect. The minimal required duty cycle $d_{\rm min}$ where the value for $\sigma_{\rm Dick}^2 (\tau)$ at optimal Ramsey time $b_2 T_{\rm opt}$ dives below the lowest variational instability is
\begin{equation}
    d_{\rm min} = \min \left\lbrace d \Big\vert \frac{b_2 T_{\rm opt}}{\chi(2) 2 \ln 2} \frac{1}{d^2} \sum_{n=1}^{\infty} \frac{\sin^2 (\pi n d)}{\pi^2 n^3} \leq \sigma^2_{\rm opt}  \right\rbrace .
\end{equation}

\begin{figure}[tb]
    \centering
    \includegraphics[width=\columnwidth]{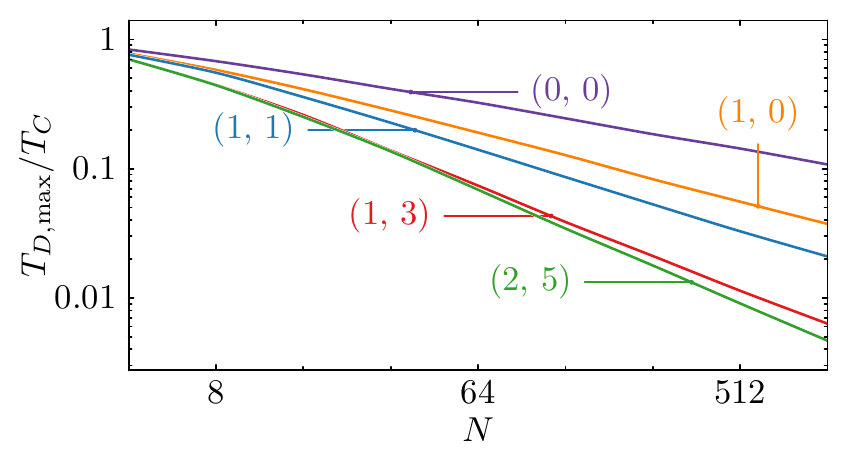}
    \caption{Largest fraction $T_{\rm D, {\rm max}}/T_{\rm C}$ of dead time compared to cycle time for which the clock operation is limited predominantly by the variationally optimized instability displayed in Fig.~\ref{fig:AllanDeviation_N}. The overall instabilities at dead time fractions above the lines are limited by the Dick effect instead.}
    \label{fig:DickEffect}
\end{figure}

From $d_{\rm min}$ one can directly infer the maximum fraction $R = T_{D, \rm max}/T_{C}= 1-T_{\rm opt}/T_{C}= 1-d_{\rm min}$ of dead time in the clock cycle, where $T_{C} = T_{\rm opt} + T_{D, \rm max}$. In the limit $R \ll 1$ it can be shown that $-\ln(R) R^2/(1-R)^2 \propto (b_2 T_{\rm opt}) \sigma_{\rm opt}^2 $, so it is expected that for $N \gg 1$ this ratio will eventually follow a similar scaling as $\sigma_{\rm opt}^2$. The exact relation is shown in Fig.~\ref{fig:DickEffect}. It is worth noting that $R \ll 1$ is still recommended for small ensemble sizes, even though this condition is not required based on $d_{\rm min}$, to prevent unnecessarily increasing the clock instability. A more complete model for the influence of dead time and the Dick effect requires to include the full spectral density $S_{L} (f)$ of the laser and evaluating the sensitivity function during the entangling and decoding dynamics.

\section{Numerical simulation of the variational clock operation}
\label{app:Clock_simulation}
In order to see how well $\sigma$ [Eq.~\eqref{eq:Allan_dimless}] reflects an achievable instability we perform numerical simulations of all essential parts involved in the closed feedback loop of an optical atomic clock when operating with the variationally optimized Ramsey protocols.

Building up the simulations proceeds as follows:

(i) The free-running laser is simulated. Given a particular spectral density $S_{L}(f) = h_{1-\alpha} f^{1- \alpha}$ and the Ramsey time $T$ we generate a sequence of random numbers $\bar{y}_k = \frac{1}{T} \int_{t_k}^{t_k+ T}dt[\omega_L (t)-\omega_A ]/\omega_A$ which gives the average frequency fluctuations of the laser without any feedback in each cycle $k$. Correlations between different cycles, required when $\alpha \neq 1$, can e.g. be obtained in the time domain by implementing $\bar{y}_k$ as a random walk or a sum of multiple damped random walks~\cite{Leroux2017}.

(ii)
To stabilise the laser frequency for long averaging times $\tau\gg  T$ a feedback correction is applied to the laser frequency at the end of each cycle. 
In the simulations, the estimated frequency deviation $\bar y_{\rm est, k} = m_k / (2 \pi \omega_A T \partial_{\phi} \bar m (\phi)_{\vert \phi =0}) $ obtained from measurement result $m_k$ at $t_k$ is multiplied by a gain factor $0 < g \leq 1$ and subtracted from the true laser frequency. 
This integrating servo corrects frequency errors over $\sim 1/g$ cycles and is sufficient to achieve a robust stabilization at $\tau/T\gg 1/g$ for flicker noise limited lasers~\cite{Schulte2020b}.
However, to simulate the quantum probabilities $p(m|\phi_k)$ at $t_k$ the phase $\phi_k = \omega_A T \bar y'_k$ based on the actual laser noise $\bar y'_k$ is needed. 
Thus, later measurements are affected not only by the noise of the free-running laser but also by the measurement results and corrections from earlier cycles.
To implement this efficiently, the simulation runs sequentially:
At the beginning the phase $\phi_1$ is calculated for the first cycle only. Then the probabilities $p(m|\phi_1)$ with this particular phase are calculated and a single measurement result $m_1$ is sampled according to this distribution. The estimator $y_{\rm est, 1}$ is calculated and the servo corrects the laser frequency so that $\bar{y}'_2 = \bar{y}_2 - g \bar y_{\rm est, 1} $ is the actual noise in the second cycle. This procedure is repeated in each cycle with the corrected frequencies, meaning e.g. $\phi_2 = \omega_A T \bar{y}'_2 $.

(iii) The clock stability is evaluated, based on the simulated sequence of stabilized frequency deviations $\bar{y}'_k$.  The overlapping Allan deviation $\sigma_y (\tau = n T)$ is calculated numerically from averages over $n$ cycles. Statistical averaging is performed over many intervals of length $n$ in a single run with $n_{\rm tot} \gg n$ cycles and then averaging again over multiple runs. Finally, the long term instability is extracted by fitting the prefactor to the asymptotic scaling $\sigma_y(\tau) \propto \tau^{1/2}$ reached typically after $n \sim 10^4$ cycles in simulations of $n_{\rm tot} = 2 \times 10^6$ cycles.

To compare numerical results to theory predictions, as in Fig.~\ref{fig:AllanDeviation_hT}(b), the values for $T$ and $h_{1-\alpha}$ in the simulations are matched to reproduce the same laser induced prior width $(\delta\phi)^2 = (b_{\alpha}T)^{\alpha}$.

\section{Cumulative interaction angle}\label{app:cum_int}
\begin{figure}[tb]
    \centering
    \includegraphics[width=\columnwidth]{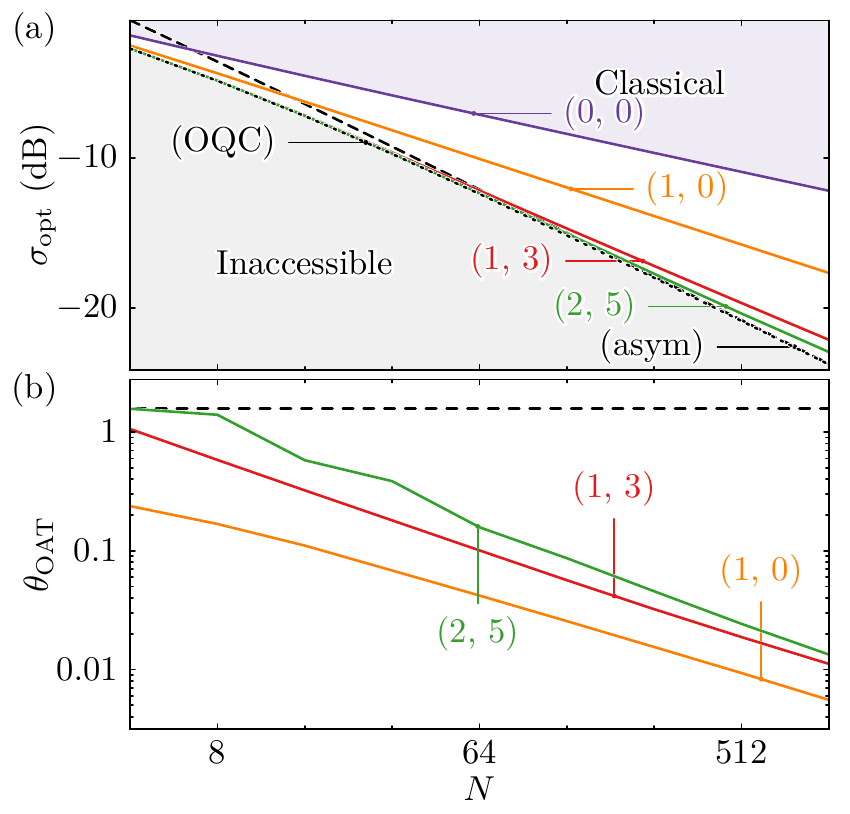}
    \caption{(a) Recalculation of the particle number dependence of the dimensionless Allan deviation in Fig.~\ref{fig:AllanDeviation_N} for constrained interaction angles. (b) The cumulative angle of all one axis twisting gates $\mathcal{T}_{x,y}$ required to obtain the dimensionless Allan deviations displayed above. The vertical dashed line indicated the interaction angle of $\pi/2$ required to prepare a GHZ-state. }
    \label{fig:CumulativeInteractionAngle}
\end{figure}

A relevant question regarding the Dick effect is the time it takes to perform the entangling and decoding sequence. The slowest time scale on a quantum simulator is usually the interaction strength. Results presented in Fig.~\ref{fig:results},~\ref{fig:AllanDeviation_hT} were obtained for interaction angles $\leq \pi / 2$. From a practical point of view, however, it might be beneficial to consider smaller interaction angles. 

Here we show that, close to the respective minima in Fig.~\ref{fig:results},~\ref{fig:AllanDeviation_hT}, the displayed results of the variationally optimized interferometers can be well approximated by quantum circuits with small cumulative interaction angles $\theta_{\rm OAT }= \sum_{k=1}^{n_{\rm En}}\big( \theta_{k}^{(1)} +  \theta_{k}^{(2)}\big) + \sum_{k=1}^{n_{\rm De}}\big( \vartheta_{k}^{(1)} +  \vartheta_{k}^{(2)}\big)$. In Fig.~\ref{fig:CumulativeInteractionAngle} we constrain each interaction angle to be positive and smaller than a threshold that decreases with the depth of the circuit. In addition we require that the cumulative interaction angle $\theta_{\rm OAT }$ is always smaller or equal than $\pi/2$, the interaction angle required to prepare a GHZ-state. Similarly to the OAT squeezing~\cite{Kitagawa1993}, the variational sequences can also work with a cumulative interaction that decrease rapidly with $N$ while the resulting Allan deviation remains a good approximation of the unconstrained optimization in Fig.~\ref{fig:AllanDeviation_N}.

\end{document}